\documentclass[twocolumn,journal]{IEEEtran}
\usepackage{cite}
\usepackage[cmex10]{amsmath}
\usepackage{array}
\usepackage{amsfonts}
\usepackage{filecontents}
\usepackage{mathrsfs}
\usepackage{arydshln}
\usepackage{slashbox}
\usepackage{graphicx}
\usepackage{float}

\usepackage{subfigure}
\usepackage{amssymb}
\usepackage{amsmath}
\usepackage{multirow}
\usepackage{multicol}
\usepackage{cite}
\usepackage[subfigure]{graphfig}
\usepackage{xcolor}
\usepackage{enumitem}
\usepackage{setspace}
\newcommand{\mv}[1]{\mbox{\boldmath{$ #1 $}}}
\usepackage[linesnumbered,ruled]{algorithm2e}

\newcommand{\qed}{\nobreak \ifvmode \relax \else
	\ifdim\lastskip<1.5em \hskip-\lastskip
	\hskip1.5em plus0em minus0.5em \fi \nobreak
	\vrule height0.75em width0.5em depth0.25em\fi}
\begin{document}
\bstctlcite{bstctl:etal, bstctl:nodash, bstctl:simpurl}
\title{{Radio Map Based 3D Path Planning for Cellular-Connected UAV}\thanks{This work has been presented in part at the IEEE Global Communications Conference (Globecom), Waikoloa, HI, USA, Dec. 9--13, 2019 \cite{GC}.}\thanks{S. Zhang is with the Department of Electronic and Information Engineering, The Hong Kong Polytechnic University (e-mail: shuowen.zhang@polyu.edu.hk). She was with the Department of Electrical and Computer Engineering, National University of Singapore. R. Zhang is with the Department of Electrical and Computer Engineering, National University of Singapore (e-mail: elezhang@nus.edu.sg).}
\author{\IEEEauthorblockN{Shuowen~Zhang, \emph{Member, IEEE} and Rui~Zhang, \emph{Fellow, IEEE}}}}
	
\maketitle
	
\begin{abstract}
In this paper, we study the three-dimensional (3D) path planning for a cellular-connected unmanned aerial vehicle (UAV) to minimize its flying distance from given initial to final locations, while ensuring a target link quality in terms of the expected signal-to-interference-plus-noise ratio (SINR) at the UAV receiver with each of its associated ground base stations (GBSs) during the flight. To exploit the location-dependent and spatially varying channel as well as interference over the 3D space, we propose a new \emph{radio map} based path planning framework for the UAV. Specifically, we consider the \emph{channel gain map} of each GBS that provides its large-scale channel gains with uniformly sampled locations on a 3D grid, which are due to static and large-size obstacles (e.g., buildings) and thus assumed to be time-invariant. Based on the channel gain maps of GBSs as well as their loading factors, we then construct an \emph{SINR map} that depicts the expected SINR levels over the sampled 3D locations.  By leveraging the obtained SINR map, we proceed to derive the optimal UAV path by solving an equivalent shortest path problem (SPP) in graph theory. We further propose a grid quantization approach where the grid points in the SINR map are more coarsely sampled by exploiting the spatial channel/interference correlation over neighboring grids. Then, we solve an approximate SPP over the reduced-size SINR map (graph) with reduced complexity. Numerical results show that the proposed solution can effectively minimize the flying distance/time of the UAV subject to its communication quality constraint, and a flexible trade-off between performance and complexity can be achieved by adjusting the grid quantization ratio in the SINR map. Moreover, the proposed solution significantly outperforms various benchmark schemes without fully exploiting the channel/interference spatial distribution in the network.
\end{abstract}
\vspace{-2mm}
\begin{IEEEkeywords}
UAV communication, cellular network, 3D path planning, radio map, graph theory.
\end{IEEEkeywords}

\vspace{-5mm}
\section{Introduction}
\vspace{-1mm}
The applications of unmanned aerial vehicles (UAVs) have become increasingly popular and diversified, ranging from cargo delivery to aerial video streaming and virtual/augmented reality \cite{cellularUAV}. To enable the safe fly of high-mobility UAVs as well as supporting timely exchange of mission data between them and their ground users, it is crucial to establish high-quality air-ground communications \cite{cellularUAV}. To this end, a promising technology is \emph{cellular-enabled UAV communication}, or cellular-connected UAV, by leveraging the ground base stations (GBSs) in the cellular network to serve the UAVs as new users in the sky \cite{cellularUAV,LTE_Sky,LinSky,Potential}. In contrast to existing technologies based on Wi-Fi over the unlicensed spectrum whose communication range is rather limited, cellular-connected UAV supports longer-range communication, and is anticipated to significantly enhance the rate and reliability performance \hbox{by leveraging the more advanced cellular technologies \cite{cellularUAV,LTE_Sky,LinSky,Potential}.}

Compared to traditional cellular communications serving only the terrestrial users, various new challenges arise in cellular-enabled UAV communications, among which two most critical issues are \emph{interference mitigation} and \emph{UAV path planning} \cite{cellularUAV}. Specifically, at high flying altitude, UAVs usually possess strong channels dominated by the \emph{line-of-sight (LoS)} path with a much larger number of GBSs compared to the terrestrial users, which leads to enhanced macro-diversity in cell association but also causes more severe \emph{co-channel interference} with terrestrial communications \cite{Network_Connected}. The strong aerial-ground interference problem calls for new and efficient interference mitigation techniques (see, e.g., \cite{multibeam,NOMA,Mei2,Mei,Cognitive,Howto}). For instance, considering multi-beam uplink transmission from a multi-antenna UAV to multiple GBSs, \cite{multibeam} proposed a novel cooperative interference cancellation technique by leveraging the backhaul connections among the GBSs, which exploits the UAV macro-diversity for cooperative processing by the GBSs for interference mitigation. Moreover, \cite{NOMA} and \cite{Mei2} devised alternative non-orthogonal multiple access (NOMA) based schemes where successive interference cancellation is performed at each GBS without the need of information exchange with the other GBSs. 

On the other hand, another appealing feature of the UAV is its high and flexible mobility in the three-dimensional (3D) space. This makes the UAV's \emph{trajectory} or \emph{path}\footnote{Note that the path and the speed along it specify the trajectory of a UAV.} design a new means for improving its communication performance with its associated GBSs via offline/online trajectory or path optimization/adaptation \cite{cellularUAV,trajectoryoutage,trajectoryoutageICC,Disconnectivity,Power_Efficient,3DMap,Interference_YW,Reinforcement}. In particular, trajectory design or path planning for cellular-connected UAV is usually performed \emph{offline} prior to the UAV's flight based on the mission requirement (e.g., flight time, initial/final locations) and available \emph{channel knowledge} with the GBSs at known locations in the UAV's fly region. For rural areas without large obstacles above the GBSs, the GBS-UAV channels can be modeled as \emph{LoS}, based on which the UAV trajectory/path optimization problems subject to various communication constraints have been studied in \cite{cellularUAV,trajectoryoutage,trajectoryoutageICC,Disconnectivity,Power_Efficient}. Specifically, \cite{cellularUAV} first investigated the trajectory optimization problem of a cellular-connected UAV for minimizing its flying time between a given pair of locations, subject to a quality-of-connectivity requirement specified by a signal-to-noise ratio (SNR) target with the associated GBSs at every time instant during the flight. By judiciously exploiting the problem structure, both the optimal solution and a polynomial-time suboptimal solution that approaches the optimal solution with arbitrarily small performance gap were proposed. Moreover, for the case with sparse GBS distribution and/or high SNR target where a feasible path that guarantees connectivity at all times may not exist, an outage cost function was proposed in \cite{trajectoryoutage} to minimize ``outage durations'' over the flight. The proposed outage cost function is general and consists of the ``maximum outage duration'' as a special case, for which the corresponding UAV trajectory design was also investigated in \cite{trajectoryoutageICC,Disconnectivity}. \cite{Power_Efficient} further studied the UAV trajectory design under connectivity constraint to minimize the propulsion energy during its flight.

\begin{figure}[t]
	\centering
	\includegraphics[width=6cm]{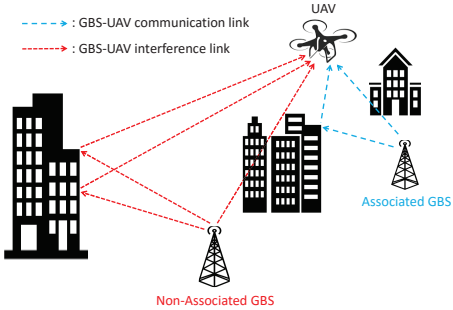}
		\vspace{-3mm}
		\caption{Cellular-enabled UAV communication under the general 3D air-ground channel model.}\label{RM_channel}
\end{figure}

Despite the rich design insights drawn from prior works, there are three major limitations of the existing studies on trajectory design for cellular-connected UAVs. Firstly, the LoS air-ground channel model is not accurate for urban/suburban environments when the UAV's altitude is not sufficiently high, where the \emph{shadowing and multi-path fading effects} become non-negligible due to signal blockage and reflection/diffraction by large-size obstacles such as buildings, as illustrated in Fig. \ref{RM_channel}. In this case, more sophisticated channel models such as \emph{Rician fading} \cite{URLLCPollin} and \emph{probabilistic LoS} \cite{LAP} models have been proposed. Based on such statistical channel models, offline UAV trajectory optimization has been studied in \cite{CoMP,You,Map_based,YW}, and a hybrid offline-online design was recently proposed in \cite{Hybrid}. It is worth noting that such statistical channel based UAV trajectory designs can only ensure the UAV communication performance in an \emph{average} sense, while the actual performance at each location along its trajectory cannot be guaranteed in general due to the lack of \emph{location-specific} channel knowledge. In this paper, we overcome this limitation via a new \emph{radio map} based approach. Generally speaking, radio maps contain rich information on the spectral activities and/or propagation channels over space and frequency in a region of interest, by averaging over the small-scale channel fading and its induced effects (e.g., power control) \cite{engineering}. In this paper, we utilize a specific type of radio maps termed as the ``\emph{channel gain map}'', which provides the large-scale channel gain between each GBS and its served UAV at any location in a given 3D region.

Secondly, the prior works \cite{cellularUAV,trajectoryoutage,trajectoryoutageICC,Disconnectivity,3DMap} have assumed that the UAV is assigned with a dedicated resource block (RB) without any aerial-ground interference. However, in practice, the UAV's RB may be reused at other non-associated GBSs even when they are distant from the UAV's serving GBS, which thus causes strong interference to/from the UAV. In this case, UAV trajectory/path design can be an effective new method for interference mitigation. For instance, in addition to flying close to the associated GBSs for enhancing the communication signal power, the UAV can also move away from the strongest interfering GBS to reduce the interference power. Nevertheless, interference power varies dynamically and spatially in practice and is difficult to be obtained at every location over the air.\footnote{It is worth noting that although the interference issue was considered in \cite{GC}, a simplified model was adopted by assuming homogeneous (worst-case) interference at all possible UAV locations, which cannot fully characterize the spatial interference distribution.} In this paper, we tackle this difficulty by leveraging the channel gain maps of GBSs together with the knowledge of the loading factor of each GBS (i.e., the average number of terrestrial users it currently serves per RB) to obtain a 3D estimation of the spatial interference distribution due to each GBS. Based on this, we construct a new ``\emph{signal-to-interference-plus-noise ratio (SINR) map}'' which depicts the expected SINR level at any location of the UAV to facilitate our proposed UAV path planning for interference mitigation, and further investigate the optimal interference-aware path design in this paper.

Thirdly, it is worth noting that the existing works \cite{GC,cellularUAV,Power_Efficient,trajectoryoutage,trajectoryoutageICC,Disconnectivity,Reinforcement,Interference_YW} have considered the two-dimensional (2D) UAV trajectory design where the UAV flies at a fixed altitude, which may be practically infeasible (e.g., when the UAV's initial and final locations have different altitudes). Moreover, under the general 3D channel models, varying the UAV's altitude may lead to further performance improvement \cite{You}. This thus motivates us to address the more general 3D path planning problem for cellular-connected UAV in this paper, based on 3D radio maps.

In summary, in this paper, we aim to develop a new \emph{radio map} based framework for designing \emph{interference-aware 3D} path for cellular-connected UAVs, under the general 3D air-ground channel model, as illustrated in Fig. \ref{RM_channel}. Our main contributions are summarized as follows.
\begin{itemize}[leftmargin=*]
	\item We consider the scenario where a cellular-connected UAV has a mission of flying between a given pair of initial and final locations, while communicating with one of the GBSs during its flight, subjected to the downlink interference from other non-associated GBSs. To minimize the mission completion time while guaranteeing satisfactory communication quality with the cellular network, we study the 3D path planning of the UAV to minimize its flying distance (or time duration with a given speed), subject to that the UAV needs to ensure a target link quality in terms of the expected SINR with each of its associated GBSs during the flight.
	\item To this end, we consider two types of radio map, namely, the ``\emph{channel gain map}'' and the newly proposed ``\emph{SINR map}''. Specifically, the channel gain map for each GBS characterizes the distribution of its large-scale channel gains over the 3D space, which can be obtained offline by deploying dedicated UAVs for channel sensing and measurements \cite{engineering,Learning,Efficient}. Particularly, efficient algorithms based on data clustering and segmented regression were proposed in \cite{Learning,Efficient} for constructing the air-ground channel gain map from small amount of measurement samples. To efficiently store and process radio maps, we assume that the channel gain maps of GBSs are obtained in practice over a finite set of uniformly sampled locations on a 3D grid. Moreover, the SINR map characterizes the expected SINR level for every sampled 3D location within the UAV's fly region, which is constructed by jointly exploiting the channel gain maps and the loading factors of different GBSs.
	\item Based on the SINR map, we transform the 3D path planning problem into an equivalent \emph{shortest path problem (SPP)} in graph theory, based on which the optimal solution is obtained via the Dijkstra algorithm \cite{graph}. Moreover, to reduce the computational complexity for finding the optimal SPP solution, we propose a grid quantization method to more coarsely sample the SINR map, by exploiting the potential spatial correlation in the channel gains/SINR levels among neighboring grid points in the maps. Then, we solve approximate SPPs over reduced-size maps/graphs to obtain suboptimal solutions with lower complexity.
	\item Last, numerical results are provided that validate the effectiveness of the proposed algorithms. Moreover, it is shown that a flexible trade-off between performance and complexity can be achieved by adjusting the map quantization ratios. In addition, compared to various benchmark schemes with imperfect/partial knowledge of the channel and/or interference spatial distribution in the network, our proposed radio map based algorithms significantly improve the UAV performance in terms of communication SINR as well as flight efficiency.
\end{itemize}

\vspace{-2mm}
It is worth noting that radio map (in particular, channel gain map) has also been considered in prior works on terrestrial communications \cite{engineering} and UAV-assisted communications with the UAV being an aerial relay/base station serving the ground users (see, e.g., \cite{Placement,Map_based}). Compared to these scenarios, radio map is more suitable for practical implementation in our considered scenario since the involved channels (i.e., GBS-UAV channels) are generally more stable due to the higher altitude of UAVs and GBSs over ground users. Moreover, compared to the UAV placement/path optimization in UAV-assisted communications based on the channel gain map \cite{Placement,Map_based}, we consider both the channel gain and SINR maps for UAV path planning as a cellular-connected user, which leads to very different problem formulations and solutions; furthermore, the dimension of the radio map in our considered scenario is generally much smaller due to the fixed GBS locations, while the radio map in the former scenario needs to store the channels between every possible pair of UAV and ground user locations. In addition, there has been a recent work \cite{coverage} that applied a similar ``aerial coverage map'' based approach for path planning of cellular-connected UAV, which, however, is limited to 2D path planning without considering the aerial-ground interference.

The remainder of this paper is organized as follows. Section \ref{sec_system} presents the system model and performance metric. Section \ref{sec_radio} introduces the SINR map construction based on the channel gain map. Section \ref{sec_problem} presents the problem formulation for the SINR-aware 3D path planning. Section \ref{sec_opt} proposes the optimal solution, while Section \ref{sec_sub} proposes a suboptimal solution with reduced complexity. Numerical examples are presented in Section \ref{sec_num}. Finally, Section \ref{sec_conclusion} concludes the paper and discusses promising directions for future work.

\emph{Notations:} Vectors and matrices are denoted by boldface lower-case letters and boldface upper-case letters, respectively. $\mv{x}^T$, $\|\mv{x}\|$, and $|\mv{x}|$ denote the transpose, the Euclidean norm, and the element-wise absolute value of a vector $\mv{x}$, respectively. For two vectors $\mv{x}$ and $\mv{y}$, ${\mv{x}}\preceq {\mv{y}}$ denotes that $\mv{x}$ is element-wise no larger than $\mv{y}$. $\mathbb{R}^{m\times n}$ denotes the space of $m\times n$ real matrices. $\mathbb{N}_+$ denotes the set of positive integers. $[\mv{X}]_{i,j,k}$ denotes the $(i,j,k)$-th element of a matrix $\mv{X}$. $\mathbb{E}[\cdot]$ denotes the statistical expectation. $|\mathcal{X}|$ denotes the cardinality of a set $\mathcal{X}$. For two sets $\mathcal{X}$ and $\mathcal{Y}$, $\mathcal{X}\backslash\mathcal{Y}$ denotes the set of elements in $\mathcal{X}$ that do not belong to $\mathcal{Y}$. $\mathcal{O}(\cdot)$ denotes the standard big-O notation. For a time-dependent function $\mv{x}(t)$, $\dot{\mv{x}}(t)$ denotes its first-order derivative with respect to time $t$.

\vspace{-4mm}
\section{System Model and Performance Metric}\label{sec_system}
\vspace{-1mm}
Consider a cellular-connected UAV and $M\geq 1$ GBSs that may potentially be associated with the UAV during its flight. The UAV has a mission of flying from an initial location $U_0$ to a final location $U_F$, while communicating with any of the $M$ GBSs during the flight. We consider a 3D Cartesian coordinate system, under which we denote ${\mv{u}}_0=[x_0,y_0,H_0]^T$ and ${\mv{u}}_F=[x_F,y_F,H_F]^T$ as the coordinates of $U_0$ and $U_F$, respectively; ${\mv{g}}_m=[a_m,b_m,H_{\mathrm{G}}]^T$ as the coordinate of each $m$th GBS, where all GBSs are assumed to have a common height $H_{\mathrm{G}}$; and ${\mv{u}}(t)=[x(t),y(t),H(t)]^T,\ 0\leq t\leq T$ as the time-varying coordinate of the UAV, with $T$ denoting the mission completion time. We assume that the UAV flies at a constant speed denoted as $V$ meter/second (m/s), thus the UAV's trajectory $\{{\mv{u}}(t):0\leq t\leq T\}$ is determined solely by its flying path. For ease of exposition, we focus on downlink transmission in this paper.

The large-scale channel gain between each GBS and the UAV constitutes the distance-dependent path loss, the shadowing, and the antenna gain, which are generally dependent on the locations of the GBS and UAV. Moreover, {small-scale fading} may also be present in the GBS-UAV channels due to random/moving scatters on the ground. Without loss of generality, let $h_m({\mv{u}})=\bar{h}_m({\mv{u}})\tilde{h}_m({\mv{u}})$ denote the instantaneous channel gain between each $m$th GBS and the UAV at location ${\mv{u}}$, where $\bar{h}_m({\mv{u}})\in \mathbb{R}$ denotes the large-scale channel gain and $\tilde{h}_m({\mv{u}})\in \mathbb{R}$ denotes the small-scale fading gain with normalized average power, i.e., $\mathbb{E}[\tilde{h}_m^2({\mv{u}})]=1$. Note that the large-scale channel gain $\bar{h}_m({\mv{u}})$ is mainly determined by the large and high obstacles (e.g., buildings) that are generally static, and thus is approximately constant for given UAV location ${\mv{u}}$ with each $m$th GBS. As such, the large-scale channel gains over different UAV locations with each GBS can be measured offline and stored in a \emph{channel gain map}, for which the details will be given in Section \ref{sec_radio}.

We assume that the UAV is associated with GBS indexed by $I(\mv{u}(t))\in \mathcal{M}$ when it is located at $\mv{u}(t)$ during its mission, over an assigned RB by its serving GBS, where $\mathcal{M}=\{1,...,M\}$. However, this RB may be reused by other GBSs for serving their terrestrial users at the same time. For each time instant $t$, let $\alpha_m(t)\in \{0,1\}$ denote the occupancy state of this RB at the $m$th GBS, with $\alpha_m(t)=1$ representing that this RB is occupied, and $\alpha_m(t)=0$ otherwise, $m\in \mathcal{M}\backslash\{I(\mv{u}(t))\}$. In practice, $\alpha_m(t)$ is determined by the real-time resource allocation of the $m$th GBS based on its channels with the associated users, thus varies in a similar time scale as $\tilde{h}_{m}({\mv{u}}(t))$. Thus, the downlink SINR at the UAV receiver can be modeled as
\begin{align}\label{SINR}
&\gamma(\mv{u}(t))=\frac{P{h}^2_{I(\mv{u}(t))}({\mv{u}}(t))}{P\sum_{m'\in \mathcal{M}\backslash \{I(\mv{u}(t))\}} \alpha_{m'}(t)h_{m'}^2({\mv{u}}(t))+\sigma^2}\nonumber\\[-1mm]
=&\frac{P\bar{h}_{I(\mv{u}(t))}^2({\mv{u}}(t))\tilde{h}^2_{I(\mv{u}(t))}({\mv{u}}(t))}{P\sum_{{m'}\in \mathcal{M}\backslash \{I(\mv{u}(t))\}} \alpha_{m'}(t)\bar{h}_{m'}^2({\mv{u}}(t))\tilde{h}^2_{m'}({\mv{u}}(t))+\sigma^2},\nonumber\\[-1mm]
& \qquad\qquad\qquad\qquad\qquad\qquad\qquad\qquad  0\leq t\leq T,
\end{align}
where $P$ denotes the transmission power of each GBS over the RB, and $\sigma^2$ denotes the noise power at the UAV receiver.

In practice, the instantaneous SINR $\gamma(\mv{u}(t))$ varies over channel coherence time, due to the fast changing $\tilde{h}_m({\mv{u}}(t))$ as well as $\alpha_m(t)$. To perform offline UAV path planning, we adopt the \emph{expected SINR} as the communication performance metric, which specifies the average quality of GBS-UAV communications; while the impairments of small-scale fading and time-varying interference can be dealt with online via countermeasures such as channel coding, power control, and dynamic RB allocation. The expected SINR is expressed as
\begin{align}\label{ExpectedSINR}
&\quad \mathbb{E}[\gamma(\mv{u}(t))]=\mathbb{E}[P\bar{h}_{I(\mv{u}(t))}^2({\mv{u}}(t))\tilde{h}^2_{I(\mv{u}(t))}({\mv{u}}(t))]\times\nonumber\\[-1mm]
& \mathbb{E}\big[1/\big(P\sum\nolimits_{{m'}\in \mathcal{M}\backslash \{I(\mv{u}(t))\}} \!\!\alpha_{m'}(t)\bar{h}_{m'}^2({\mv{u}}(t))\tilde{h}^2_{m'}({\mv{u}}(t))\!+\!\sigma^2\big)\big]\nonumber\\[-1mm]
&\ =P\bar{h}_{I(\mv{u}(t))}^2({\mv{u}}(t))\times\nonumber\\[-1mm]
&\mathbb{E}\big[1/\big(P\sum\nolimits_{{m'}\in \mathcal{M}\backslash \{I(\mv{u}(t))\}} \!\!\alpha_{m'}(t)\bar{h}_{m'}^2({\mv{u}}(t))\tilde{h}^2_{m'}({\mv{u}}(t))\!+\!\sigma^2\big)\big],\nonumber\\[-1mm]
& \qquad\qquad\qquad\qquad\qquad\qquad\qquad\qquad 0\leq t\leq T.
\end{align}
Note that the expected SINR is determined by the distributions of two sets of random variables, namely, $\alpha_m(t)$'s and $\tilde{h}_{m}({\mv{u}}(t))$'s. In practice, $\alpha_m(t)$ can be modeled as a Bernoulli random variable with mean $l_m$, namely, $\mathbb{E}[\alpha_m(t)]=l_m$, where $l_m\in [0,1]$ denotes the the so-called \emph{loading factor} for each $m$th GBS. Specifically, $l_m$ represents the probability of the RB assigned for the UAV being occupied at the $m$th GBS, which can be roughly estimated as the average number of users served by each GBS over the total number of available RBs, over a given period of time. As each $l_m$ varies slowly in practice, $\{l_m\}_{m=1}^M$ can be obtained/updated efficiently in the network. In contrast, distributions of the small-scale fading gains $\tilde{h}_{m}({\mv{u}}(t))$'s vary over time and space more rapidly, which may not be available in practice. Moreover, it is generally difficult to express the expected SINR in (\ref{ExpectedSINR}) in a tractable form even if such distributions are known. Therefore, we approximate the expected SINR by its lower bound given below
\begin{align}\label{bound}
& \mathbb{E}[\gamma(\mv{u}(t))]\overset{(a)}{\geq}\nonumber\\[-1mm]
&\frac{P\bar{h}_{I(\mv{u}(t))}^2({\mv{u}}(t))}{\mathbb{E}\left[P\sum_{m'\in \mathcal{M}\backslash \{I(\mv{u}(t))\}} \alpha_{m'}(t)\bar{h}_{m'}^2({\mv{u}}(t))\tilde{h}^2_{m'}({\mv{u}}(t))+\sigma^2\right]}\nonumber\\[-1mm]
=&\frac{P\bar{h}_{I(\mv{u}(t))}^2({\mv{u}}(t))}{P\sum_{{m'}\in \mathcal{M}\backslash \{I(\mv{u}(t))\}} l_{m'}\bar{h}_{m'}^2({\mv{u}}(t))+\sigma^2},\ 0\leq t\leq T,
\end{align}where $(a)$ holds due to the Jensen's inequality since the function $\frac{1}{x}$ is convex over $x$ for $x>0$. To maximize the (approximate) expected SINR in (\ref{bound}), the associated GBS with the UAV should be selected as 
\vspace{-2mm}\begin{align}
I(\mv{u}(t))=&\arg\underset{m\in \mathcal{M}}{\max}\ \frac{P\bar{h}_m^2({\mv{u}}(t))}{P\sum_{{m'}\in \mathcal{M}\backslash \{m\}} l_{m'}\bar{h}_{m'}^2({\mv{u}}(t))+\sigma^2},\nonumber\\[-1mm]
&\qquad\qquad\qquad\qquad\qquad\qquad\qquad 0\leq t\leq T.
\vspace{-2mm}\end{align} 
Consequently, the expected SINR with the associated GBS is given by
\begin{align}\label{channelconstraint}
\bar{\gamma}({\mv{u}}(t))\overset{\Delta}{=}\underset{m\in \mathcal{M}}{\max}\ \frac{P\bar{h}_m^2({\mv{u}}(t))}{P\sum_{{m'}\in \mathcal{M}\backslash \{m\}} l_{m'}\bar{h}_{m'}^2({\mv{u}}(t))+\sigma^2},\nonumber\\[-1mm]
 0\leq t\leq T.
\vspace{-1mm}\end{align}
Note that the expected SINR given in (\ref{channelconstraint}) is only determined by the large-scale channel gains and loading factors, thus can be calculated offline efficiently.

In the next, we first characterize the large-scale channel gains $\bar{h}_m(\mv{u}(t))$'s between every UAV location ${\mv{u}}(t)$ and GBS $m$ based on its channel gain map, then construct a so-called \emph{SINR map} that characterizes the expected SINR at every UAV location ${\mv{u}}(t)$, i.e., $\bar{\gamma}({\mv{u}}(t))$ defined in (\ref{channelconstraint}), based on the channel gain maps and the loading factors $\{l_m\}_{m=1}^M$. Furthermore, based on the SINR map, we formulate and solve the UAV 3D path planning problem {\hbox{under a constraint on the expected SINR.}}

\begin{figure*}[t]
	\centering
	\subfigure[GBS and obstacle locations]{
		\includegraphics[width=6.5cm]{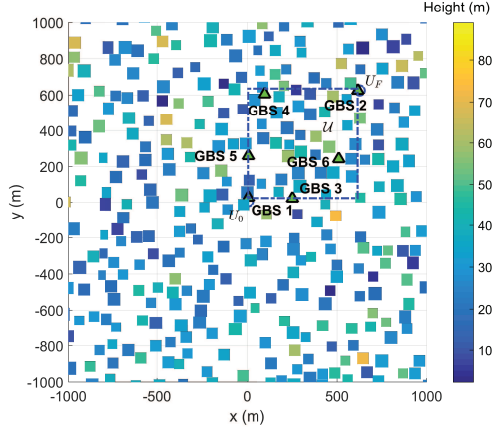}}
	\subfigure[Channel gain map for GBS 6 at $H=125$ m]{
		\includegraphics[width=6.5cm]{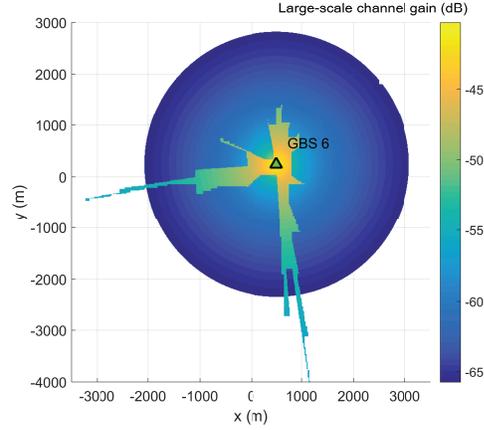}}
	\vspace{-3mm}
	\caption{Illustration of the channel gain map.}\label{Radio_Map}
	\vspace{-6mm}
\end{figure*}

\vspace{-3mm}
\section{SINR Map Construction}\label{sec_radio}
\vspace{-1mm}
In this section, we aim to construct the SINR map needed for the UAV path planning. Specifically, we first introduce the detailed structure and storage of the channel gain map as mentioned in Section \ref{sec_system}. Based on this, we then present the construction of the SINR map.
\vspace{-3mm}
\subsection{Channel Gain Map}\label{sec_channelmap}
\vspace{-1mm}
First, we introduce the channel gain map. The channel gain map for each $m$th GBS refers to the spatial distribution of its large-scale channel gain over the 3D space, i.e., $\bar{h}_m({\mv{u}})$'s with UAV at locations ${\mv{u}}\in \mathbb{R}^{3\times 1}$. As the space is infinite and continuous, it is not feasible to store the entire data $\{\bar{h}_m({\mv{u}}):{\mv{u}}\in \mathbb{R}^{3\times 1}\}$ for all locations of $\mv{u}$, due to the finite storage in practice.

To achieve efficient storage, the channel gain map of each $m$th GBS can be depicted only for a \emph{truncated} 3D space consisting of only neighborhood locations with \emph{non-negligible} large-scale channel gains above a given threshold, denoted by $\epsilon$, so as to reduce the map size. Moreover, the space can be \emph{discretized} into a 3D grid with a finite granularity $\Delta_D$, where $\Delta_D$ is chosen to be sufficiently small such that the channel gain is approximately constant within each grid cell. Thus, the channel gain map of each $m$th GBS can be efficiently represented by a 3D \emph{matrix} of finite size denoted by ${\bar{\mv{H}}}_m\in \mathbb{R}^{{X}_m\times {Y}_m\times {Z}_m}$, in which each element represents the large-scale channel gain between the $m$th GBS and each corresponding location in the truncated and discretized 3D space, denoted as ${\mathcal{U}}_D^m=\{{\mv{u}}_D^m(i^m,j^m,k^m)\in \mathbb{R}^{3\times 1}:\ i^m\in \mathcal{X}_m,j^m\in \mathcal{Y}_m,k^m\in \mathcal{Z}_m\}$, with $\mathcal{X}_m=\{1,...,X_m\}$, $\mathcal{Y}_m=\{1,...,Y_m\}$, $\mathcal{Z}_m=\{1,...,Z_m\}$. Specifically, each element in ${\bar{\mv{H}}}_m$ is given by
\vspace{-1mm}\begin{align}
[{\bar{\mv{H}}}_m]_{i^m,j^m,k^m}=&\bar{h}_m({\mv{u}}_D^m(i^m,j^m,k^m)),\nonumber\\
&\qquad\quad i^m\in \mathcal{X}_m, j^m\in \mathcal{Y}_m, k^m\in \mathcal{Z}_m.
\vspace{-1mm}\end{align}
Note that the size of ${\bar{\mv{H}}}_m$ as well as ${\mathcal{U}}_D^m$ specified by ${X}_m$, ${Y}_m$, and ${Z}_m$ is determined by the number of discretized 3D locations that yield large-scale channel gains no smaller than $\epsilon$; while for simplicity, we assume $\bar{h}_m({\mv{u}})=0$ for ${\mv{u}}$ that is outside the locations considered in ${\mathcal{U}}_D^m$. The location of each $(i^m,j^m,k^m)$-th element in ${\mathcal{U}}_D^m$ can be further expressed as
\vspace{-1mm}\begin{align}\label{uDm}
{\mv{u}}_D^m(i^m,j^m,k^m)={\mv{u}}_{R}^m+\left[i^m-1,j^m-1,k^m-1\right]^T\Delta_D,\nonumber\\
 i^m\in \mathcal{X}_m, j^m\in \mathcal{Y}_m, k^m\in \mathcal{Z}_m,
\vspace{-1mm}\end{align}
where ${\mv{u}}_{R}^m\in \mathbb{R}^{3\times 1}$ denotes the reference location in ${\mathcal{U}}_D^m$ with the smallest coordinates over all three dimensions.\footnote{For ease of storage, we assume that the locations in each ${\mathcal{U}}_D^m$ form a regular 3D grid without loss of generality.} Therefore, based on knowledge of the channel gain map ${\bar{\mv{H}}}_m$, we can obtain the large-scale channel gain between GBS $m$ and any UAV location over the 3D space. Note that for each GBS, its channel gain map needs to store only $X_mY_mZ_m+4$ real numbers (i.e., the elements in $\bar{\mv{H}}_m$, $\mv{u}_R^m$, and $\Delta_D$). As an example, for an area with GBS and obstacle locations given in Fig. \ref{Radio_Map} (a), where $H_{\mathrm{G}}=10$ m, Fig. \ref{Radio_Map} (b) shows the channel gain map for GBS 6 at altitude $H=125$ m, with $\epsilon=-65.724$ dB (which is chosen such that the average received power from the GBS is less than the noise power if $\bar{h}_m({\mv{u}})<\epsilon$) and $\Delta_D=10$ m, while the other parameters will be given later in Section \ref{sec_num}. It can be observed that in such a dense urban environment, the large-scale channel gain behaves differently from that under the LoS channel model, where two different locations with equal distance to the same GBS may have drastically different gains due to heterogeneous shadowing effects. In addition, the number of non-zero grid points in the channel gain map at $H=125$ m is no larger than $3\times 10^5$, thanks to the proposed map truncation and discretization method.

In the sequel of this paper, we assume that the channel gain maps for the $M$ GBSs, $\{\bar{\mv{H}}_m\}_{m=1}^M$, are \emph{perfectly} known with granularity $\Delta_D$, which is \emph{sufficiently small} such that $\bar{h}_m({\mv{u}})=[\bar{\mv{H}}_m]_{i,j,k}$ holds for any UAV location ${\mv{u}}$ in the $(i,j,k)$-th grid \emph{cell} (i.e., ${\mv{u}}$ that satisfies $|{\mv{u}}-{\mv{u}}^m_D(i,j,k)|\preceq\frac{\Delta_D}{2}[1,1,1]^T$).

\begin{figure}[t]
	\centering
	\includegraphics[width=4.5cm]{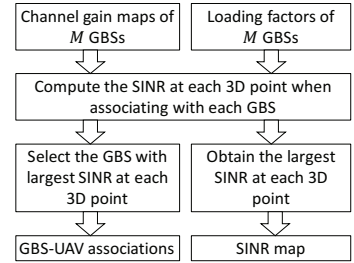}
	\vspace{-3mm}
	\caption{Flowchart for SINR map construction.}\label{Chart}
\end{figure}
\begin{figure*}[t]
	\centering
	\subfigure[GBS-UAV associations over $\mathcal{U}$ at $H=125$ m]{
		\includegraphics[width=6.5cm]{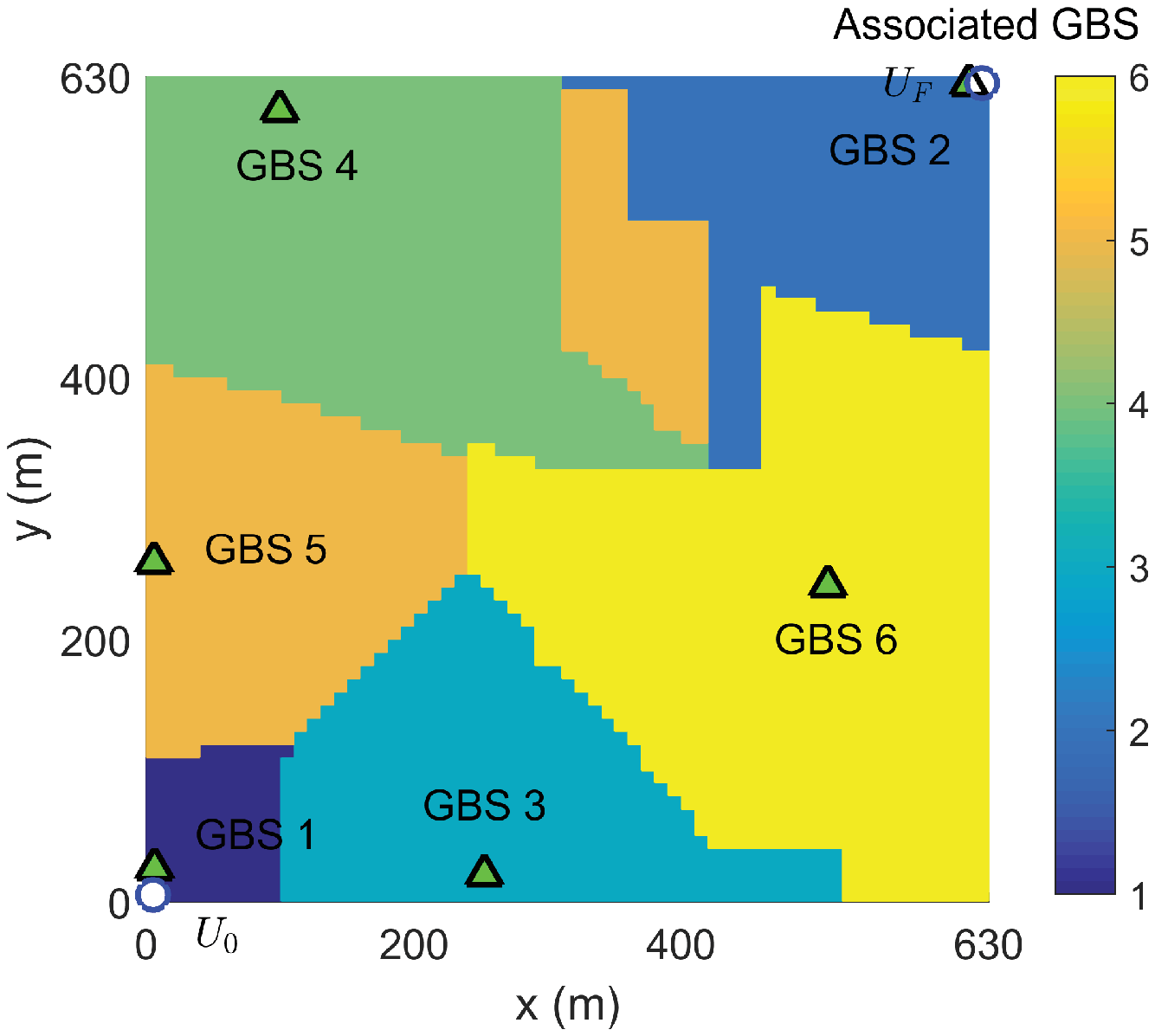}}
	\subfigure[SINR map over $\mathcal{U}$ at $H=125$ m]{
		\includegraphics[width=6.5cm]{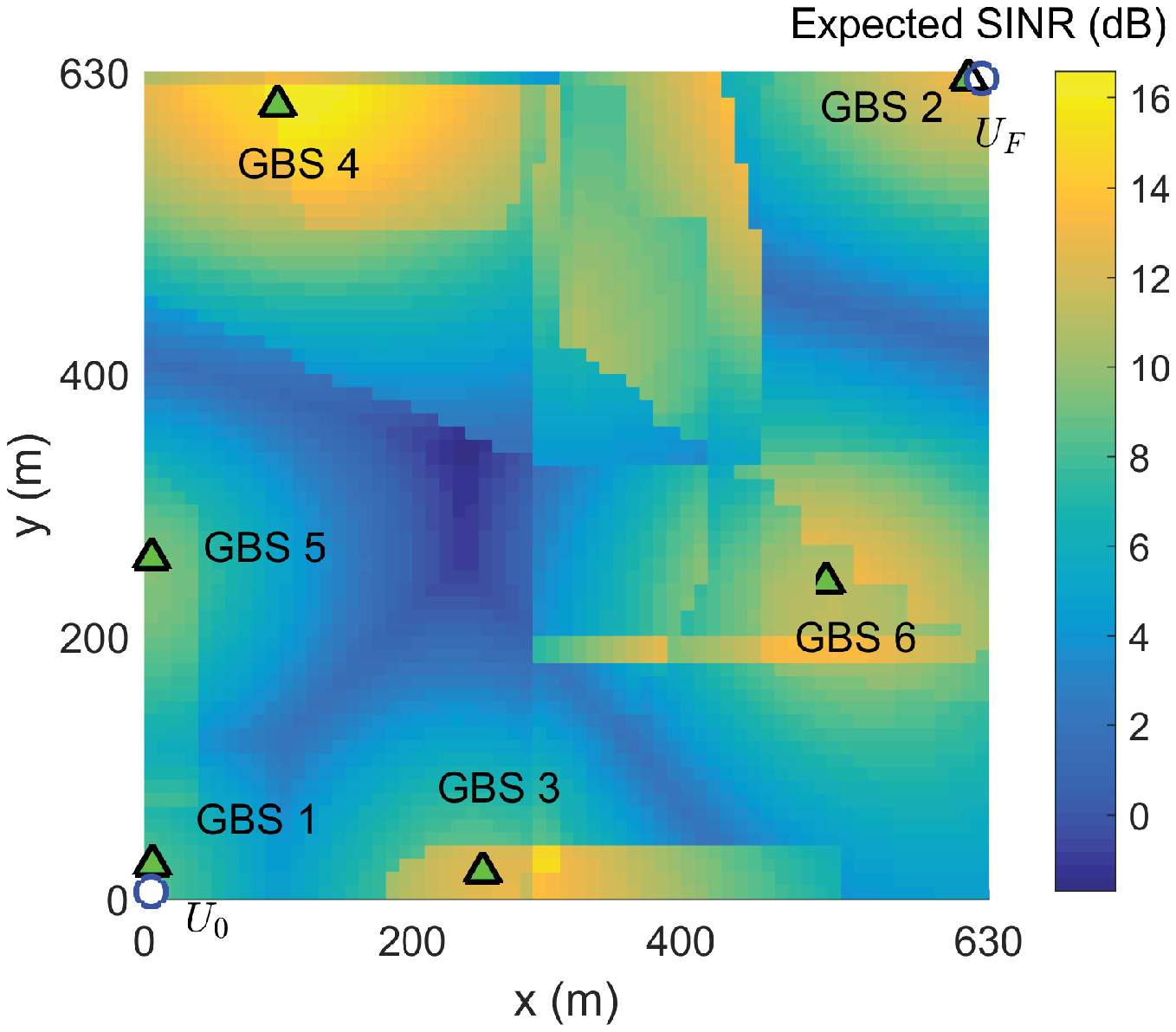}}
	\vspace{-2mm}
	\caption{Illustration of the GBS-UAV associations and the SINR map.}\label{SINR_Map}
	\vspace{-6mm}
\end{figure*}

\subsection{SINR Map}
\vspace{-2mm}
Next, we construct the SINR map based on the channel gain maps $\{\bar{\mv{H}}_m\}_{m=1}^M$ and loading factors $\{l_m\}_{m=1}^M$ of the $M$ GBSs, to infer the expected SINR defined in (\ref{channelconstraint}) at any UAV location. For UAV path planning with given initial and final locations, we only need to consider the SINR map constructed by the channel gain maps for the $M$ GBSs that overlap with a target region which is sufficiently large to cover all possible UAV locations during its flight. Specifically, the UAV's flying altitude is typically constrained as $H(t)\in [H_{\min},H_{\max}],\ \forall t\in [0,T]$, where $H_{\min}$ denotes the minimum allowable altitude to avoid collisions with the ground obstacles, and $H_{\max}$ denotes the maximum allowable altitude specified by government regulations. Moreover, the horizontal location of the UAV can be assumed to be constrained in a square region denoted by $\mathcal{U}_{\mathrm{H}}\subset \mathbb{R}^{2\times 1}$ with edge length $L$, which is chosen to be sufficiently large to cover all possible UAV horizontal locations during the flight. Note that $L$ generally increases with the horizontal distance between $U_0$ and $U_F$. As such, we only need to consider the UAV's locations during its flight in a 3D square cuboid region denoted by ${\mathcal{U}}\subset \mathbb{R}^{3\times 1}$, with length $L$ and height $H_{\mathrm{R}}\overset{\Delta}{=}H_{\max}-H_{\min}$. Considering the same discretization granularity $\Delta_D$ as the channel gain maps, the discretized UAV locations in ${\mathcal{U}}$ form a $D\times D\times Z$ grid, where $D=L/\Delta_D$ and $Z=H_{\mathrm{R}}/\Delta_D$. Such a grid can be represented by ${\mathcal{U}}_D=\{{\mv{u}}_D(i,j,k)\in \mathbb{R}^{3\times 1}:i,j\in {\mathcal{D}},k\in \mathcal{Z}\}$, with $\mathcal{D}=\{1,...,D\}$, $\mathcal{Z}=\{1,...,Z\}$, and ${\mv{u}}_D(i,j,k)$ \hbox{denoting the $(i,j,k)$-th location on the grid, which is given by}
\vspace{-2mm}\begin{equation}
{\mv{u}}_D(i,j,k)\!=\!\Big[i\!-\!\frac{1}{2},j\!-\!\frac{1}{2},k\!-\!\frac{1}{2}\Big]^T\Delta_{{D}},i,j\in {\mathcal{D}},k\in \mathcal{Z}.
\vspace{-2mm}\end{equation}

Based on the above, our objective is to construct an SINR map that characterizes the expected SINR at every UAV location in ${\mathcal{U}}_D$, which can be represented by a 3D matrix $\bar{\mv{S}}\in \mathbb{R}^{D\times D\times Z}$. Specifically, each $(i,j,k)$-th element in $\bar{\mv{S}}$ is given by
\vspace{-2mm}\begin{align}\label{SINRmap}
[\bar{\mv{S}}]_{i,j,k}=&\underset{m\in \mathcal{M}}{\max}\ \frac{P\bar{h}^2_m({\mv{u}}_D(i,j,k))}{\tilde{\sigma}^2({\mv{u}}_D(i,j,k))-Pl_m\bar{h}^2_m({\mv{u}}_D(i,j,k))},\nonumber\\ 
&\qquad\qquad\qquad\qquad i,j\in \mathcal{D},k\in \mathcal{Z},
\vspace{-2mm}\end{align}
where $\tilde{\sigma}^2({\mv{u}}_D(i,j,k))\overset{\Delta}{=}\sigma^2+P\sum_{{m'}\in \mathcal{M}} l_{m'}\bar{h}_{m'}^2({\mv{u}}_D(i,j,k))$. Thus, obtaining $[\bar{\mv{S}}]_{i,j,k}$ requires the extraction of the large-scale channel gain between ${\mv{u}}_D(i,j,k)$ and all the $M$ GBSs, i.e., $\{\bar{h}_{m}({\mv{u}}_D(i,j,k))\}_{m=1}^M$. According to the channel gain map storage method in Section \ref{sec_channelmap}, each $\bar{h}_m({\mv{u}}_D(i,j,k))$ can be extracted from $\bar{\mv{H}}_m$ based on its parameters ${\mv{u}}_R^m$ and $\Delta_D$ as
\begin{align}\label{map_matrix}
&\bar{h}_m({\mv{u}}_D(i,j,k))=\nonumber\\
&\begin{cases}[\bar{\mv{H}}_m]_{i^m,j^m,k^m},\nonumber\\
\qquad\qquad \mathrm{if}\ |{\mv{u}}_D(i,j,k)\!-\!{\mv{u}}_D^m(i^m,j^m,k^m)|\!\preceq\! \frac{\Delta_D}{2}[1,1,1]^T,\nonumber\\[-1mm]
\qquad\qquad\qquad\qquad i^m\in \mathcal{X}_m,j^m\in \mathcal{Y}_m,k^m\in \mathcal{Z}_m\\[-1mm]
0, \qquad\quad  \mathrm{otherwise}  \end{cases}\nonumber\\[-1mm]
&\quad\qquad\qquad\qquad\qquad\qquad m\in \mathcal{M},\ i,j\in \mathcal{D}, k\in \mathcal{Z},
\end{align}
where ${\mv{u}}_D^m(i^m,j^m,k^m)$ is defined by ${\mv{u}}_R^m$ and $\Delta_D$ in (\ref{uDm}). Particularly, if ${\mv{u}}_D(i,j,k)$ does not belong to the effective channel gain map of the $m$th GBS, its corresponding large-scale channel gain is set as zero; otherwise, it is set as the large-scale channel gain of its belonged cell in ${\mathcal{U}}_D^m$. Then, for each $(i,j,k)$-th location in ${\mathcal{U}}_D$, we need to first calculate $\tilde{\sigma}^2({\mv{u}}_D(i,j,k))$, based on which we can further obtain $\frac{P\bar{h}^2_m({\mv{u}}_D(i,j,k))}{\tilde{\sigma}^2({\mv{u}}_D(i,j,k))-Pl_m\bar{h}^2_m({\mv{u}}_D(i,j,k))}$ for all $m\in \mathcal{M}$, and select the optimal associated GBS with the maximum expected SINR as given in (\ref{channelconstraint}). The overall complexity of the aforementioned procedure over all locations in ${\mathcal{U}}_D$ can be shown to be $\mathcal{O}(D^2ZM)$. In Fig. \ref{Chart}, we provide a flowchart to summarize the SINR map construction procedure based on the channel gain maps and GBS loading factors.

For illustration, with an example of $\mathcal{U}$ given in Fig. \ref{Radio_Map} (a), we illustrate in Fig. \ref{SINR_Map} (a) the GBS-UAV associations over $\mathcal{U}$ at $H=125$ m, for a given set of loading factors which will be specified in Section \ref{sec_num}. It is observed that under the location-specific channel and interference, the GBS associated with the UAV at any location is not necessarily the one with the smallest distance to the UAV, which is in sharp contrast to the case of LoS channels considered in prior works \cite{cellularUAV,trajectoryoutage,trajectoryoutageICC,Disconnectivity}. Moreover, we show in Fig. \ref{SINR_Map} (b) the SINR map over $\mathcal{U}$ at $H=125$ m. It is observed that the SINR map varies more abruptly than the channel gain map of individual GBSs shown in Fig. \ref{Radio_Map} (b), since the SINR is a function of the channel gains and interference powers over multiple GBSs.

\vspace{-3mm}
\section{Problem Formulation for SINR-Aware 3D Path Planning}\label{sec_problem}
	\vspace{-1mm}
Based on the given SINR map (i.e., $\bar{\mv{S}}$ in (\ref{SINRmap})), we aim to minimize the UAV's flying time/distance from $U_0$ to $U_F$ by optimizing its 3D path denoted by $\{{\mv{u}}(t):\|\dot{{\mv{u}}}(t)\|= V, 0\leq t\leq T\}$, subject to a constraint on the expected SINR along the UAV path, with $V$ denoting the constant speed of the UAV. Specifically, we consider in this paper a given target on the expected SINR denoted by $\bar{\gamma}_{\mathrm{T}}$, which needs to be achieved throughout the UAV's flight to meet the minimum link quality required for its mission (e.g., for receiving the command and control signal from its associated GBS).\footnote{It is worth noting that besides a given SINR target that needs to be achieved throughout the flight,  various other functions of the SINR may also be considered in the constraint, e.g., by specifying a threshold on the maximum continuous flying distance or the total flying distance that the SINR target is not achieved (see, e.g., \cite{trajectoryoutage,trajectoryoutageICC}). In Section \ref{sec_num}, we will briefly discuss the extension of our algorithm to such new constraints.} Based on (\ref{bound})--(\ref{channelconstraint}), this can be achieved if
\vspace{-1mm}\begin{equation}\label{target}
\bar{\gamma}(\mv{u}(t))\geq \bar{\gamma}_{\mathrm{T}},\quad 0\leq t\leq T.
\vspace{-1mm}\end{equation}
This optimization problem is thus formulated as
\vspace{-1mm}\begin{align}
\mbox{(P0)}\quad \underset{T,\{{\mv{u}}(t),0\leq t\leq T\}}{\min} &T\\[-1mm]
\mathrm{s.t.}\qquad &\bar{\gamma}({\mv{u}}(t))\geq \bar{\gamma}_{\mathrm{T}},\quad 0\leq t\leq T\\[-1mm]
& {\mv{u}}(0)={\mv{u}}_0\\[-1mm]
& {\mv{u}}(T)={\mv{u}}_F\\[-1mm]
& \|\dot{{\mv{u}}}(t)\|=V,\quad 0\leq t\leq T.
\end{align}

Note that the above continuous path planning problem involves an infinite number of optimization variables, which are generally difficult to handle. Moreover, unlike our prior work \cite{cellularUAV} under the LoS channel model with no interference where the optimal path structure can be shown to simplify the problem, the expected SINR constraint considered in this paper is dependent on both the location-specific large-scale channel gain and interference power, which thus makes (P0) more challenging to solve.

\begin{figure}[t]
\centering
	\includegraphics[width=8cm]{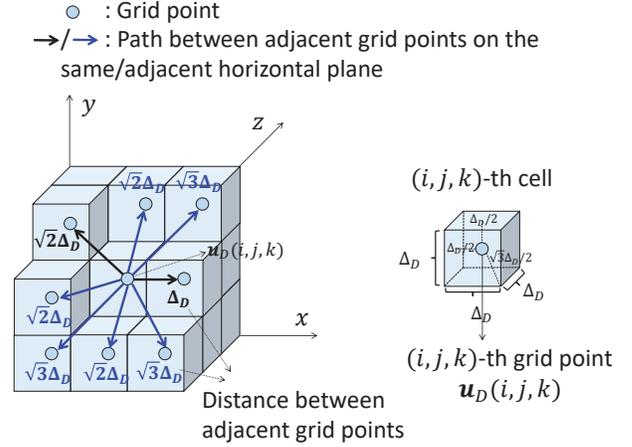}
	\vspace{-2mm}
	\caption{Illustration of 3D grid and path structure for (P1).}\label{grid1}
\end{figure}

Nevertheless, we can show that the optimal path solution to (P0) is \emph{piecewise linear}, where each waypoint (except the first and last ones as the initial and final UAV locations, respectively) lies in the intersection of two \emph{adjacent} cells that both satisfy the expected SINR target. This can be proved following the similar procedure as that for Proposition 3 in \cite{cellularUAV} by noting that each cell is a \emph{convex polyhedron}, thus sharing similar properties as the convex-shaped GBS coverage area in \cite{cellularUAV}. Thus, we omit the proof here for brevity. To relieve the burden of finding the optimal waypoint locations under this structure, we consider an approximate path structure composed of \emph{connected line segments}, where two end points of each segment are \emph{adjacent grid points} from ${\mathcal{U}}_D$ with distance no larger than  $\sqrt{3}\Delta_D$, as illustrated in Fig. \ref{grid1}. This is motivated by the fact that if two adjacent grid points both satisfy the expected SINR constraint, any point ${\mv{u}}$ on the line segment between them also satisfies this constraint, since the grid granularity $\Delta_D$ is chosen to be sufficiently small such that all points within each cell have the same channel/SINR level, and all points on the line segment between two adjacent grid points are guaranteed to lie within the two cells. For convenience, we assume that the initial and final locations of the UAV, ${\mv{u}}_0$ and ${\mv{u}}_F$, are both on the grid ${\mathcal{U}}_D$. Therefore, we formulate the following discrete path planning problem over the 3D grid:
\begin{align}
\mbox{(P1)}\!\!\!
\underset{\scriptstyle N,\atop \scriptstyle \{i_n,j_n,k_n\}_{n=1}^N}{\min}\!\! &\!\!\sum_{n=1}^{N-1}\!\! \|{\mv{u}}_{D}(i_{n+1},j_{n+1},k_{n+1})\!-\!{\mv{u}}_D(i_n,j_n,k_n)\|\!\!\\[-1mm]
\mathrm{s.t.}\qquad &[\bar{{\mv{S}}}]_{i_n,j_n,k_n}\geq \bar{\gamma}_{\mathrm{T}},\quad n=1,...,N\label{P1c1}\\[-1mm]
& {\mv{u}}_D(i_1,j_1,k_1)={\mv{u}}_0\label{P1c2}\\[-1mm]
& \|{\mv{u}}_D(i_{n+1},j_{n+1},k_{n+1})-{\mv{u}}_D(i_n,j_n,k_n)\|\nonumber\\
&\qquad \quad \leq \sqrt{3}{\Delta}_D,\quad n=1,...,N-1\\[-1mm]
& {\mv{u}}_D(i_N,j_N,k_N)={\mv{u}}_F\label{P1c4}\\[-1mm]
& i_n,j_n\in \mathcal{D},\quad n=1,...,N\\[-1mm]
& k_n\in \mathcal{Z},\quad n=1,...,N,
\end{align}
where $N$ denotes the number of grid points that the UAV traverses over its flight. Note that (P1) is a non-convex combinatorial optimization problem due to the integer variables $\{i_n,j_n,k_n\}_{n=1}^N$ and $N$. Thus, it cannot be solved efficiently via standard optimization methods. In the following, we reformulate (P1) based on graph theory, and propose both the optimal and suboptimal solutions for it.

\vspace{-4mm}
\section{Optimal Solution}\label{sec_opt}
\vspace{-2mm}
In this section, we obtain the optimal solution to (P1) by recasting it as an equivalent \emph{SPP} in graph theory \cite{graph}. To this end, a straightforward approach is to consider all the $D^2Z$ grid points in ${\mathcal{U}}_D$ in the vertex set of an equivalent graph. However, this may be inefficient since under the constraints in (\ref{P1c1}), only the $(i,j,k)$-th grid points with $[\bar{{\mv{S}}}]_{i,j,k}\geq\bar{\gamma}_{\mathrm{T}}$ may potentially constitute a feasible path. Thus, we first consider the following \emph{radio map preprocessing} to identify such valid grid points, which are referred to as the \emph{``feasible grid points''}.

\vspace{-4mm}
\subsection{Radio Map Preprocessing}
\vspace{-1mm}
Specifically, we construct a new \emph{``feasible map''} denoted by ${\mv{F}}\in \{0,1\}^{D\times D\times Z}$ based on $\bar{\mv{S}}$, where each $(i,j,k)$-th element is given by
\vspace{-2mm}\begin{equation}\label{feasible}
[\mv{F}]_{i,j,k}=\begin{cases}
1,\ \mathrm{if}\ [\bar{\mv{S}}]_{i,j,k}\geq \bar{\gamma}_{\mathrm{T}}\\
0,\ \mathrm{otherwise},
\end{cases} i,j\in \mathcal{D},k\in \mathcal{Z}.
\vspace{-2mm}\end{equation}
Note that $[\mv{F}]_{i,j,k}=1$ indicates that the $(i,j,k)$-th grid point is a feasible grid point, and $[\mv{F}]_{i,j,k}=0$ otherwise. The complexity for the above preprocessing can be shown to be $\mathcal{O}(D^2Z)$.
\vspace{-5mm}
\subsection{Graph Based Problem Reformulation and Solution}
\vspace{-1mm}
Next, based on the constructed feasible map ${\mv{F}}$, we propose an equivalent graph based reformulation of (P1). Specifically, we construct an undirected weighted graph $G_{\mathrm{D}}=(V_{\mathrm{D}},E_{\mathrm{D}})$ \cite{graph}. The vertex set of $G_{\mathrm{D}}$ is given by
\vspace{-2mm}\begin{equation}
V_{\mathrm{D}}=\{U_D(i,j,k):[{\mv{F}}]_{i,j,k}=1,i,j\in \mathcal{D},k\in \mathcal{Z}\},
\vspace{-2mm}\end{equation}
where $\!U_{{D}}(i,j,k)\!$ represents the $(i,j,k)$-th (feasible) grid point with location $\!{\mv{u}}_D(i,j,k)\!$. The edge set of $G_{\mathrm{D}}$ is given by
\vspace{-2mm}\begin{align}
E_{\mathrm{D}}=&\{(U_D(i,j,k),U_D(i',j',k')):\nonumber\\[-1mm]
&\|{\mv{u}}_D(i,j,k)-{\mv{u}}_D(i',j',k')\|\leq \sqrt{3}{\Delta}_D\}.
\vspace{-2mm}\end{align}
Note that an edge exists between two vertices $\!U_D(i,j,k)\!$ and $\!U_D(i',j',k')\!$ if and only if the corresponding two grid points are adjacent. Moreover, the weight of each edge is given by
\vspace{-2mm}\begin{equation}
W_{\mathrm{D}}(U_D(i,j,k),U_D(i',j',k'))\!=\!\|{\mv{u}}_D(i,j,k)\!-\!{\mv{u}}_D(i',j',k')\|,
\vspace{-1mm}\end{equation}
which represents the flying distance between the two corresponding locations.

Prior to solving (P1), we first check its feasibility based on $G_{\mathrm{D}}$. Specifically, it can be shown that (P1) is feasible if and only if $U_D(i_1,j_1,k_1)$ and $U_D(i_N,j_N,k_N)$ are \emph{connected} in $G_{\mathrm{D}}$, where $(i_1,j_1,k_1)$ and $(i_N,j_N,k_N)$ are given in (\ref{P1c2}) and (\ref{P1c4}), respectively. Such connectivity can be checked via various existing algorithms such as the breadth-first search (BFS) \cite{graph} with complexity $\mathcal{O}(|V_{\mathrm{D}}|+|E_{\mathrm{D}}|)=\mathcal{O}(D^2Z)$, where the worst-case values for $|E_{\mathrm{D}}|$ and $|V_{\mathrm{D}}|$ can be shown to be $2(D-1)(2D-1)(3Z-2)+D^2(Z-1)\propto D^2Z$ and $D^2Z$, respectively. It can also be shown that constructing the graph $G_{\mathrm{D}}$ requires worst-case complexity of $\mathcal{O}(D^2Z)$, thus the overall worst-case complexity for checking the feasibility of (P1) is $\mathcal{O}(D^2Z)$. Note that as the SINR target increases, the feasible map and graph $G_{\mathrm{D}}$ will become more sparse, and it is more likely for (P1) to be infeasible. If (P1) is verified to be feasible, with graph $G_{\mathrm{D}}$ constructed above, (P1) can be shown to be equivalent to finding the \emph{shortest path} from $U_D(i_1,j_1,k_1)$ to $U_D(i_N,j_N,k_N)$ in $G_{\mathrm{D}}$. This problem can be solved via the Dijkstra algorithm with worst-case complexity $\mathcal{O}(|E_{\mathrm{D}}|+|V_{\mathrm{D}}|\log|V_{\mathrm{D}}|)=\mathcal{O}(D^2Z\log (D^2Z))$ using the Fibonacci heap structure \cite{graph}. With the obtained shortest path denoted by $(U_D(i_1^\star,j_1^\star,k_1^\star),...,U_D(i_{N^\star}^\star,j_{N^\star}^\star,k_{N^\star}^\star))$, the optimal solution to (P1) is obtained as $N^\star$ and $\{i_n^\star,j_n^\star,k_n^\star\}_{n=1}^{N^\star}$.

\vspace{-3mm}
\section{Suboptimal Solution via Grid Quantization}\label{sec_sub}
\vspace{-1mm}
Note that the \emph{complexity} for finding the optimal solution to (P1) scales up with $D$ and $Z$, while the values of $D=\frac{L}{\Delta_D}$ and $Z=\frac{H_{\mathrm{R}}}{\Delta_D}$ can be practically arbitrarily large with given $\Delta_D$ and increasing the edge length $L$ of the horizontal fly region of interest, $\mathcal{U}_{\mathrm{H}}$, and/or the allowable UAV altitude range $H_{\mathrm{R}}$. Moreover, it is worth noting that the required \emph{memory} for storing all the edge weights in graph $G_{\mathrm{D}}$ is dependent on $|E_{\mathrm{D}}|$, which also increases with $D$ and $Z$. For example, with $\Delta_D=10$ m, $L=200$ km when $U_0$ and $U_F$ are far apart, and $H_{\mathrm{R}}=40$ m, we have $D=2\times 10^4$, $Z=4$, and consequently $D^2Z\log (D^2Z)\approx 3.4\times 10^{10}$; in addition, we have $|E_{\mathrm{D}}|\approx 1.72\times 10^{10}$ in the worst case, which demands for approximately $137.6$ GB memory for storing $G_{\mathrm{D}}$. In practice, such high complexity and large memory size are prohibitive or even unaffordable. To tackle this issue, we propose to reduce the number of vertices involved in the SPP by applying a \emph{grid quantization method}, and considering a new path structure composed of connected line segments between \emph{quantized grid points}, as detailed next.

\vspace{-5mm}
\subsection{Radio Map Preprocessing}
\begin{figure}[t]
	\centering
	\includegraphics[width=9cm]{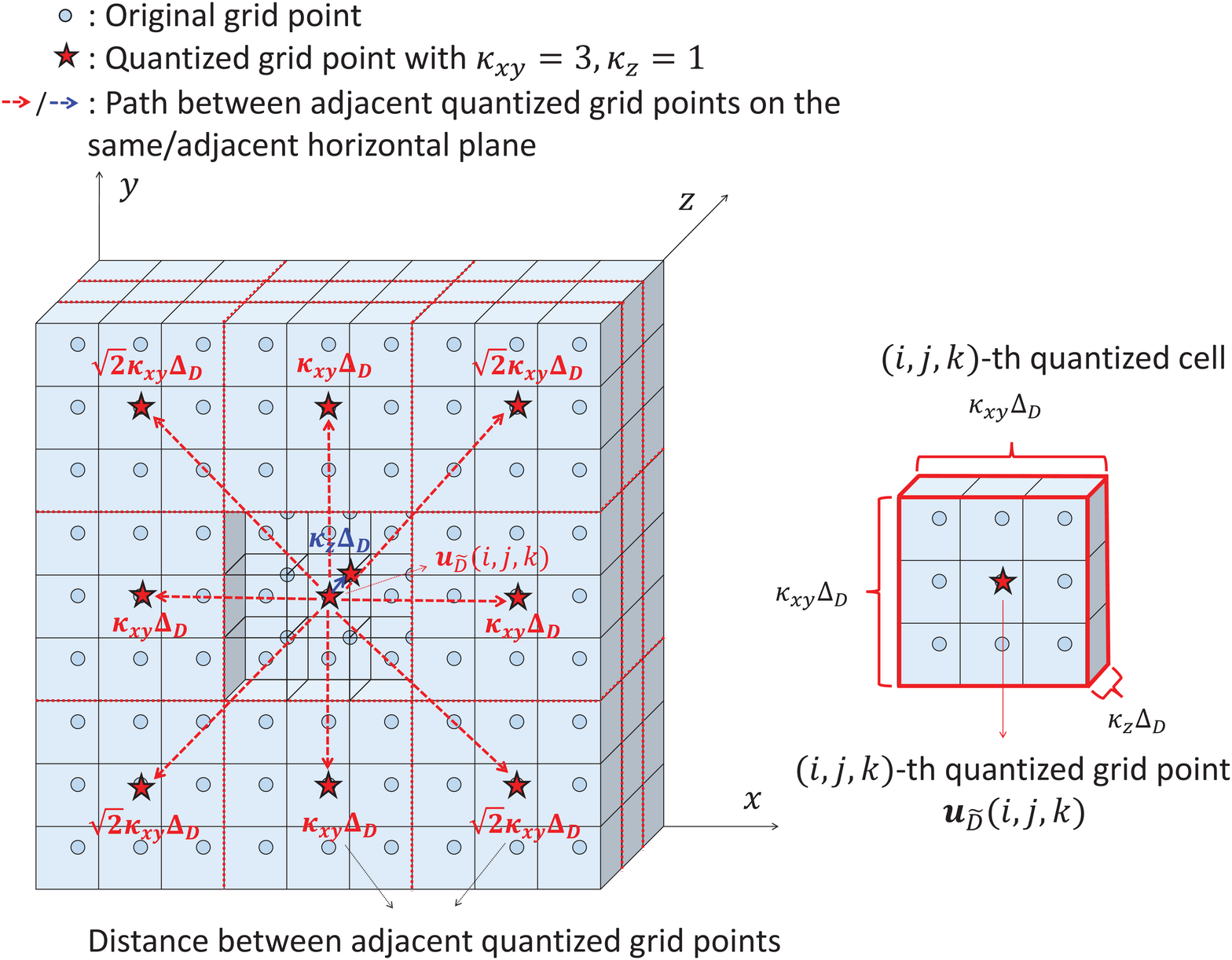}
		\vspace{-3mm}
		\caption{Illustration of the proposed grid quantization method and suboptimal path structure for (P1).}\label{quantization}
		\vspace{-1mm}
\end{figure}

\vspace{-1mm}
To start with, we present the proposed grid quantization method. Note that in practice, the allowable UAV altitude range $H_{\mathrm{R}}$ is typically much smaller than the edge length $L$ of the horizontal fly region of the UAV (as shown in the above example). Therefore, we propose to perform larger level of quantization for the horizontal dimensions (i.e., $x-y$ axes as illustrated in Figs. \ref{SINR_Map}--\ref{grid1}) and smaller level of quantization for the vertical dimension (i.e., $z$ axis as illustrated in Fig. \ref{grid1}). Specifically, let $\kappa_{xy}\in \mathbb{N}_+$ and $\kappa_z\in \mathbb{N}_+$ denote the \emph{horizontal quantization ratio} and \emph{vertical quantization ratio}, respectively, with $\kappa_{xy}\geq 1$, $\kappa_z\geq 1$, and $\kappa_{xy}\geq \kappa_z$. By applying \emph{uniform quantization} over the grid points in $\mathcal{U}_D$ according to the given quantization ratios, we obtain $\tilde{D}^2\tilde{Z}$ points with horizontal granularity $\Delta_{\tilde{D}_{xy}}$ and vertical granularity $\Delta_{\tilde{D}_z}$, where $\tilde{D}=D/\kappa_{xy}\leq D$, $\Delta_{\tilde{D}_{xy}}=\kappa_{xy}\Delta_D\geq \Delta_D$; $\tilde{Z}=Z/\kappa_{z}\leq Z$, $\Delta_{\tilde{D}_{z}}=\kappa_{z}\Delta_D\geq \Delta_D$. For ease of exposition, we assume that $D/\kappa_{xy}$ and $Z/\kappa_{z}$ are both integers, and $\kappa_{xy}$ and $\kappa_z$ are both odd numbers. Let $\mathcal{U}_{\tilde{D}}=\{{\mv{u}}_{\tilde{D}}(i,j,k):i,j\in \tilde{\mathcal{D}},k\in \tilde{\mathcal{Z}}\}$ denote the \emph{quantized grid}, with $\tilde{\mathcal{D}}=\{1,...,\tilde{D}\}$, $\tilde{\mathcal{Z}}=\{1,...,\tilde{Z}\}$, and ${\mv{u}}_{\tilde{D}}(i,j,k)$ denoting the $(i,j,k)$-th location on the quantized grid, which is given by
\vspace{-1mm}\begin{align}
{\mv{u}}_{\tilde{D}}(i,j,k)\!=&\!\left[\left(i\!-\!\frac{1}{2}\right)\Delta_{\tilde{D}_{xy}},\left(j\!-\!\frac{1}{2}\right)\Delta_{\tilde{D}_{xy}},\left(k\!-\!\frac{1}{2}\right)\Delta_{\tilde{D}_z}\right]^T\!\!\!\!,\nonumber\\
&\qquad\qquad\qquad\qquad i,j\in\tilde{\mathcal{D}},k\in \tilde{\mathcal{Z}}.
\end{align}
The proposed grid quantization method is illustrated in Fig. \ref{quantization} for the case of $\kappa_{xy}=3$ and $\kappa_z=1$. Notice that each ${\mv{u}}_{\tilde{D}}(i,j,k)$ lies among $\kappa_{xy}^2\kappa_z$ original grid points indexed by a \emph{``neighboring set''} $\mathcal{N}(i,j,k)$, whose corresponding cells form a \emph{``quantized cell''}, as illustrated in Fig. \ref{quantization}. Specifically, we have $\mathcal{N}(i,j,k)=\{(p,q,l):|{\mv{u}}_D(p,q,l)-{\mv{u}}_{\tilde{D}}(i,j,k)|\preceq \frac{1}{2}[\Delta_{\tilde{D}_{xy}},\Delta_{\tilde{D}_{xy}},\Delta_{\tilde{D}_{z}}]^T,p,q\in \mathcal{D},l\in \mathcal{Z}\}$. This is motivated by the fact that the channels and consequently SINR values for neighboring grid points in $\mathcal{N}(i,j,k)$ are typically highly correlated, thus they can be ``well-represented'' by one single quantized grid point ${\mv{u}}_{\tilde{D}}(i,j,k)$ at the center.

Next, we consider a feasible path structure where ${\mv{u}}_{\tilde{D}}(i,j,k)$ is a \emph{``feasible quantized grid point''} (which may potentially constitute a feasible path) if and only if all the neighboring original grid points in $\mathcal{N}(i,j,k)$ are the feasible grid points defined by (\ref{feasible}), such that its connected line segment with another adjacent feasible quantized grid point at any direction does not violate the expected SINR constraint. To identify such points, we construct a new \emph{``quantized feasible map''} denoted by $\tilde{\mv{F}}\in \{0,1\}^{\tilde{D}\times \tilde{D}\times \tilde{Z}}$, where each $(i,j,k)$-th element \hbox{is given by}
\vspace{-1mm}\begin{align}
[\tilde{\mv{F}}]_{i,j,k}=&\begin{cases}
1,\ \mathrm{if}\ [\bar{\mv{S}}]_{p,q,l}\geq \bar{\gamma}_{\mathrm{T}}, \forall (p,q,l)\in \mathcal{N}(i,j,k)\\[-1mm]
0,\ \mathrm{otherwise},
\end{cases}\nonumber\\[-2mm]
&\qquad\qquad\qquad\qquad i,j\in \tilde{\mathcal{D}},k\in \tilde{\mathcal{Z}}.
\end{align}
Note that $[\tilde{\mv{F}}]_{i,j,k}=1$ indicates that the $(i,j,k)$-th quantized grid point is a feasible quantized grid point, and $[\tilde{\mv{F}}]_{i,j,k}=0$ otherwise. The complexity for the above preprocessing can be shown to be $\mathcal{O}(\tilde{D}^2\tilde{Z}\kappa_{xy}^2\kappa_z)=\mathcal{O}(D^2Z)$.

\vspace{-6mm}
\subsection{Reduced-Size Graph and Suboptimal Solution}
\begin{figure}[t]
	\centering
	\includegraphics[width=9cm]{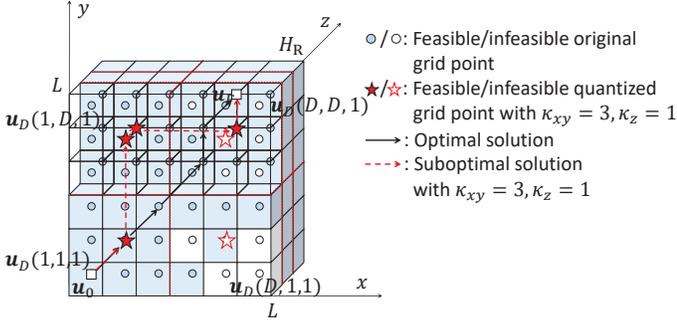}
	\vspace{-3mm}
	\caption{Illustration of proposed optimal and suboptimal path solutions for (P1).}\label{grid}
	\vspace{-1mm}
\end{figure}

\vspace{-1mm}
Based on the quantized feasible map $\tilde{\mv{F}}$, we can introduce the corresponding new path structure. Specifically, assuming that ${\mv{u}}_0$ and ${\mv{u}}_F$ belong to the $(\tilde{i}_1,\tilde{j}_1,\tilde{k}_1)$-th and $(\tilde{i}_N,\tilde{j}_N,\tilde{k}_N)$-th quantized cells, respectively. We first let the UAV fly from ${\mv{u}}_0$ to ${\mv{u}}_{\tilde{D}}(\tilde{i}_1,\tilde{j}_1,\tilde{k}_1)$ at the start, and from ${\mv{u}}_{\tilde{D}}(\tilde{i}_N,\tilde{j}_N,\tilde{k}_N)$ to ${\mv{u}}_F$ in the end. Moreover, we assume that a path exists between two feasible quantized grid points if and only if they are adjacent, and this path does not pass through any quantized cell other than the two corresponding ones. Thus, the UAV can fly from a feasible quantized grid point indexed by $(i,j,k)$ to adjacent points in $10$ directions, including $8$ on the same horizontal plane with distance $\kappa_{xy}\Delta_D$ or $\sqrt{2}\kappa_{xy}\Delta_D$, as well as $2$ on the adjacent horizontal planes with distance $\kappa_z\Delta_D$, whose index set is given by $\mathcal{A}(i,j,k)=\{(i',j',k'):\|{\mv{u}}_{\tilde{D}}(i,j,k)-{\mv{u}}_{\tilde{D}}(i',j',k')\|\in \{\kappa_{xy}\Delta_D,\sqrt{2}\kappa_{xy}\Delta_D,\kappa_{z}\Delta_D\},i',j'\in \tilde{\mathcal{D}},k'\in \tilde{\mathcal{Z}}\}$, as illustrated in Fig. \ref{quantization}.\footnote{Note that due to the heterogeneous horizontal versus vertical quantization ratios, only $10$ out of the $26$ paths between a quantized grid point and its $26$ 3D neighbors are guaranteed to lie within the two corresponding quantized cells.} Under the above structure, we construct an undirected weighted graph $G_{\tilde{\mathrm{D}}}=(V_{\tilde{\mathrm{D}}},E_{\tilde{\mathrm{D}}})$ with vertex set
\vspace{-2mm}\begin{equation}
V_{\tilde{\mathrm{D}}}=\{U_{\tilde{D}}(i,j,k):[\tilde{\mv{F}}]_{i,j,k}=1,i,j\in \tilde{\mathcal{D}},k\in \tilde{\mathcal{Z}}\},
\vspace{-2mm}\end{equation}
where $U_{\tilde{D}}(i,j,k)$ denotes the $(i,j,k)$-th (feasible) quantized grid point with location ${\mv{u}}_{\tilde{D}}(i,j,k)$. Note that $|V_{\tilde{\mathrm{D}}}|$ is significantly smaller than $|V_{{\mathrm{D}}}|$, with a worst-case value of $\tilde{D}^2\tilde{Z}=D^2Z/(\kappa_{xy}^2\kappa_z)\leq D^2Z$. The edge set of $G_{\tilde{\mathrm{D}}}$ is then given by
\vspace{-2mm}\begin{equation}\label{edgeI}
E_{\tilde{\mathrm{D}}}=\{(U_{\tilde{D}}(i,j,k),U_{\tilde{D}}(i',j',k')):(i',j',k')\in \mathcal{A}(i,j,k)\}.
\vspace{-2mm}\end{equation}
The weight of each edge is given by
\vspace{-2mm}\begin{equation}
W_{\tilde{\mathrm{D}}}(U_{\tilde{D}}(i,j,k),U_{\tilde{D}}(i',j',k'))=\|{\mv{u}}_{\tilde{D}}(i,j,k)-{\mv{u}}_{\tilde{D}}(i',j',k')\|.
\vspace{-1mm}\end{equation}
Note that (P1) under the above path structure is equivalent to finding the \emph{shortest path} from $U_{\tilde{D}}(\tilde{i}_1,\tilde{j}_1,\tilde{k}_1)$ to $U_{\tilde{D}}(\tilde{i}_N,\tilde{j}_N,\tilde{k}_N)$ in graph $G_{\tilde{\mathrm{D}}}$, which can be solved via the Dijkstra algorithm with worst-case complexity $\mathcal{O}(|E_{\tilde{\mathrm{D}}}|+|V_{\tilde{\mathrm{D}}}|\log|V_{\tilde{\mathrm{D}}}|)=\mathcal{O}(\tilde{D}^2\tilde{Z}\log (\tilde{D}^2\tilde{Z}))$ \cite{graph}. By noting that the construction of graph $G_{\tilde{\mathrm{D}}}$ requires the worst-case complexity of $\mathcal{O}(\tilde{D}^2\tilde{Z})$, the overall worst-case complexity for obtaining a subptimal solution based on $\tilde{\mv{F}}$ is $\mathcal{O}(\tilde{D}^2\tilde{Z}\log (\tilde{D}^2\tilde{Z}))=\mathcal{O}\left(\frac{D^2Z}{\kappa_{xy}^2\kappa_z}\log \left(\frac{D^2Z}{\kappa_{xy}^2\kappa_z}\right)\right)$. Recall the example discussed at the beginning of this section, with $\kappa_{xy}=25$ and $\kappa_z=1$, we have $\tilde{D}^2\tilde{Z}\log (\tilde{D}^2\tilde{Z})\approx 5.4\times 10^7$ and $|E_{\tilde{\mathrm{D}}}|\approx 2.75\times 10^7$ in the worst case. Thus, only $0.22$ GB memory is required for storing the reduced graph $G_{\tilde{\mathrm{D}}}$, which is more affordable in practice. For illustration, an example of the proposed suboptimal path solution is given in Fig. \ref{grid} with $\kappa_{xy}=3$ and $\kappa_z=1$, as compared to the optimal path solution obtained in Section \ref{sec_opt}. It can be easily shown that the obtained shortest path in $G_{\tilde{\mathrm{D}}}$ always corresponds to a feasible solution to (P1) (as can be observed from Fig. \ref{grid}), which is optimal to (P1) with $\kappa_{xy}=1$ and $\kappa_z=1$ (i.e., $\tilde{D}=D$, $\tilde{Z}=Z$), and generally suboptimal for $\kappa_{xy}>1$ and/or $\kappa_z>1$.

Finally, note that the overall complexities for the proposed optimal solution and suboptimal solution are given by $\mathcal{O}\left(D^2Z\!+\!D^2Z\log (D^2Z)\right)$ and $\mathcal{O}\left(D^2Z\!+\!\frac{D^2Z}{\kappa_{xy}^2\kappa_z}\log \left(\frac{D^2Z}{\kappa_{xy}^2\kappa_z}\right)\right)$, respectively, which can be well-approximated by $\mathcal{O}(D^2Z\log (D^2Z))$ and $\mathcal{O}\left(\frac{1}{\kappa_{xy}^2\kappa_z}D^2Z\log (D^2Z)\right)$, respectively, for the practical case with $D\gg \kappa_{xy}$ and $Z\gg \kappa_z$. Thus, the suboptimal solution only requires $1/(\kappa_{xy}^2\kappa_z)$ of the complexity required by the optimal solution. Note that as $\kappa_{xy}$ and/or $\kappa_z$ increases, the performance of the suboptimal solution generally degrades as the quantization becomes more coarse, while the required complexity also decreases. Thus, a flexible performance-complexity trade-off can be achieved by selecting the quantization ratios $\kappa_{xy}$ and $\kappa_z$.

\vspace{-3mm}
\section{Numerical Examples}\label{sec_num}
\vspace{-1mm}
In this section, we provide numerical examples to evaluate the performance of our proposed 3D path planning algorithms for UAV. We set the minimum and maximum flying altitude of the UAV as $H_{\min}\!=\!90$ m and $H_{\max}\!=\!130$ m, respectively, which correspond to an allowable UAV altitude range of $H_{\mathrm{R}}\!=\!H_{\max}\!-\!H_{\min}\!=\!40$ m. The UAV's initial and final locations are set as ${\mv{u}}_0\!=\![5,5,95]^T$ m and ${\mv{u}}_F\!=\![625,625,125]^T$ m, respectively. As illustrated in Fig. \ref{Radio_Map} (a), we consider a square horizontal area $\mathcal{U}_{\mathrm{H}}$ with edge length $L\!=\!630$ m, over which $M\!=\!6$ GBSs are uniformly randomly distributed, each with height $H_{\mathrm{G}}\!=\!10$ m under the urban micro (UMi) setup \cite{3GPPUAV}; moreover, $30$ obstacles are randomly distributed in $\mathcal{U}_{\mathrm{H}}$, each modeled as a 3D cuboid with equal length and width randomly generated according to the uniform distribution in $[50,70]$ m, and height randomly generated according to the Rayleigh distribution with mean $30$ m \cite{LAP}, which is truncated to be no larger than the UAV's minimum flying altitude $H_{\min}\!=\!90$ m. We consider an overall bandwidth of $10$ MHz, over which there are $50$ RBs, each with bandwidth $180$ KHz \cite{3GPPUAV}. The total transmission power at each GBS is set as $41$ dBm, thus that over each RB is $P\!=\!24.0103$ dBm \cite{3GPPUAV}. The noise power spectrum density is set as $-169$ dBm/Hz, with a $9$ dB noise figure \cite{3GPPUAV}. We assume that the UAV is equipped with an isotropic antenna with unit gain. The large-scale channel gain between each GBS and UAV location is modeled according to the 3GPP technical report based on the UMi scenario \cite{3GPPUAV}. Specifically, a GBS-UAV channel is classified as an \emph{LoS channel} if there is no obstacle between the GBS and the UAV, which is modeled by
\begin{align}\label{channel_LoS}
&\bar{h}_m^{\mathrm{LoS}}({\mv{u}})=\frac{G_m^A(\mv{u})}{2}+
\frac{1}{2}\min\{\bar{h}_m^{\mathrm{FS}},-30.9-(22.25\nonumber\\
&\qquad-0.5\log_{10}H(\mv{u}))\log_{10}d_m({\mv{u}})-20\log_{10}f_c\},
\end{align}
in dB, with $G_m^A(\mv{u})$ denoting the GBS antenna radiation power gain at UAV location $\mv{u}$ in dB, $\bar{h}_m^{\mathrm{FS}}$ denoting the free-space path loss, $d_m({\mv{u}})\overset{\Delta}{=}\|{\mv{u}}-{\mv{g}}_m\|$ denoting the 3D GBS-UAV distance, $H(\mv{u})$ denoting the UAV's altitude at location $\mv{u}$, and $f_c=2$ GHz denoting the carrier frequency; otherwise, it is classified as an \emph{NLoS (non-LoS) channel} modeled by 
\begin{align}\label{channel_NLoS}
&\bar{h}_m^{\mathrm{NLoS}}({\mv{u}})=\frac{G_m^A(\mv{u})}{2}+
\frac{1}{2}\min\{\bar{h}_m^{\mathrm{LoS}}({\mv{u}}), -32.4-(43.2\nonumber\\
&\qquad-7.6\log_{10}H({\mv{u}}))\log_{10}d_m({\mv{u}})-20\log_{10}f_c\},\!\!
\end{align}
in dB. Note that we have $\bar{h}_m^{\mathrm{LoS}}({\mv{u}})\geq \bar{h}_m^{\mathrm{NLoS}}({\mv{u}})$ since the path loss of NLoS channel is generally larger than its LoS counterpart, thus resulting in smaller channel gain. For the purpose of exposition, we first consider in Section \ref{sec_num}-A to Section \ref{sec_num}-C the simplified scenario where each GBS has an isotropic gain, i.e., $G_m^A(\mv{u})=0$ dB. The corresponding radio maps are illustrated in Fig. \ref{Radio_Map} and Fig. \ref{SINR_Map}. Then, in Section \ref{sec_num}-D, we consider the GBS antenna model in the current LTE standard, where each GBS is equipped with a uniform linear array (ULA) consisting of $8$ antenna elements with half-wavelength spacings, which are electronically tilted downwards with elevation angle $\theta_{\mathrm{tilt}}=-10^{\mathrm{o}}$. The synthesized antenna gain between each GBS and UAV location, $G_m^A(\mv{u})$, is specified in \cite{3GPPmodel}. Unless stated otherwise, we consider a set of loading factors given by $\mv{l}=[l_1,...,l_M]^T=[0.0318,0.6561,0.3223,0.9679,0.2598,0.7672]^T\overset{\Delta}{=}\bar{\mv{l}}$, each of which is independently and randomly generated according to uniform distribution in $[0,1]$. The granularity for the channel gain maps and the SINR map is set as $\Delta_D=10$ m.
\vspace{-3mm}
\subsection{Performance of Proposed Solution with Different Quantization Ratios}
\begin{figure}[t]
	\centering
	\subfigure[$\bar{\gamma}_{\mathrm{T}}=0$ dB]{
		\includegraphics[width=9cm]{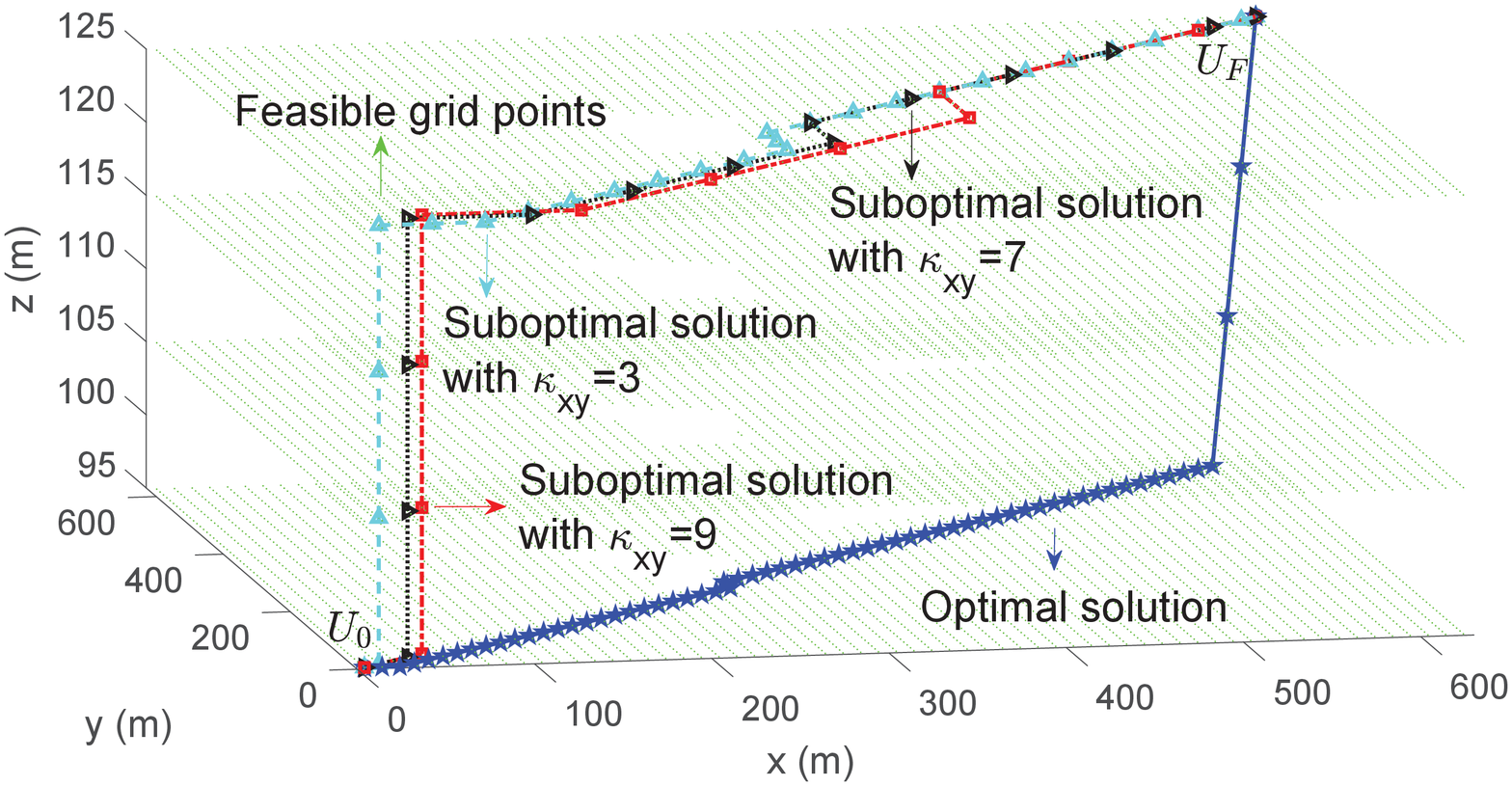}}
	\subfigure[$\bar{\gamma}_{\mathrm{T}}=3$ dB]{
		\includegraphics[width=9cm]{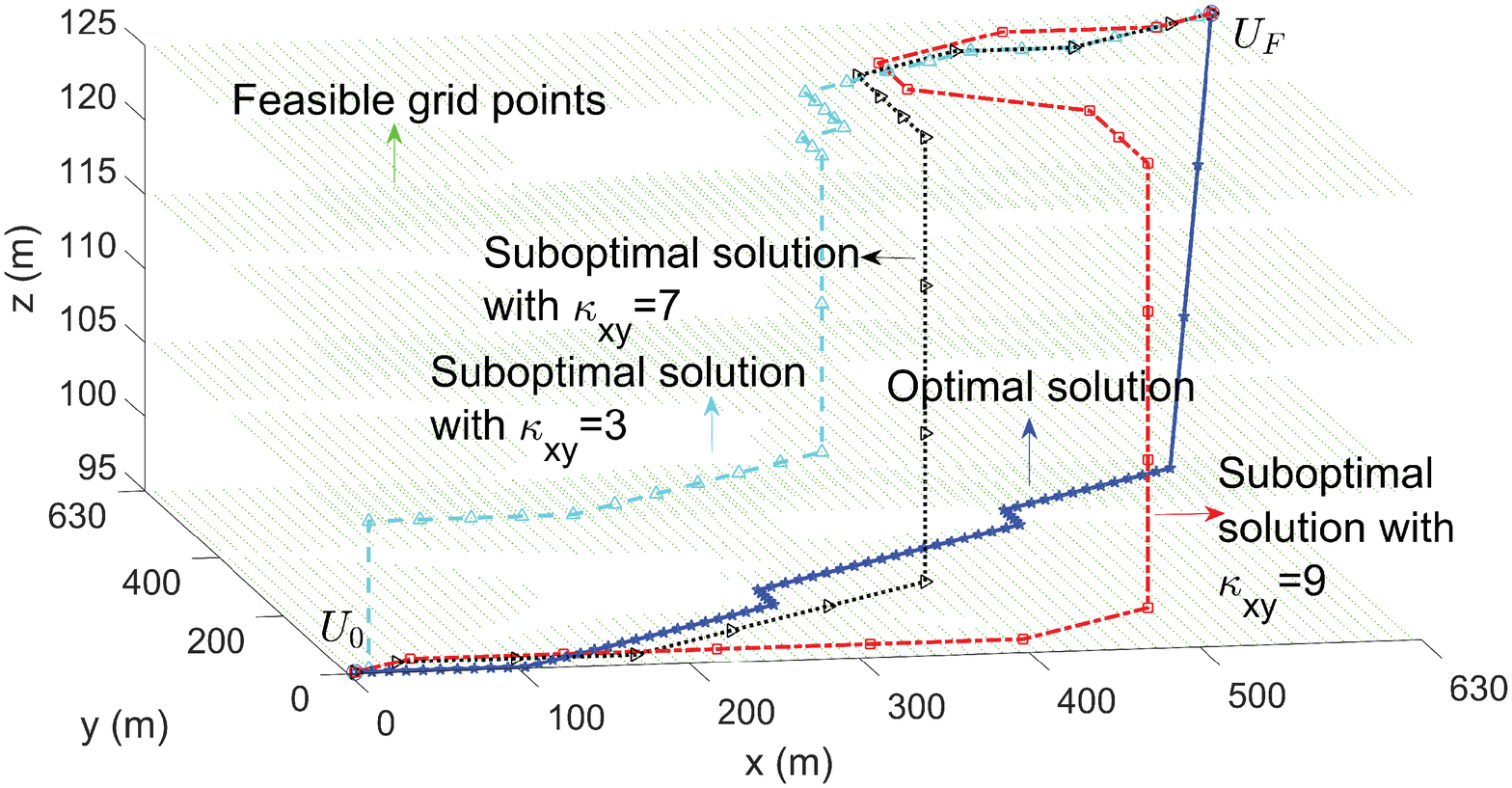}}
	\vspace{-3mm}
	\caption{Illustration of the proposed path solutions with different quantization ratios.}\label{path}
		\vspace{-3mm}
\end{figure}

To start with, we evaluate the efficacy of our proposed grid quantization method by comparing the performance of our proposed optimal solution and suboptimal solutions with different quantization ratios. Specifically, since $L$ is much larger than $H_{\mathrm{R}}$, we fix the vertical quantization ratio as $\kappa_z=1$, and consider a set of different horizontal quantization ratios given as  $\kappa_{xy}\in \{3,7,9\}$. Under this setup, we consider two expected SINR targets in the downlink, $\bar{\gamma}_{\mathrm{T}}=0$ dB or $3$ dB, and show in Fig. \ref{path} the proposed optimal and suboptimal path solutions with the three horizontal quantization ratios. For the purpose of illustration, we also depict in Fig. \ref{path} the feasible (original) grid points that satisfy the expected SINR target. For both expected SINR targets, it is observed that the suboptimal solutions with different quantization levels $\kappa_{xy}$ as well as the optimal solution are substantially different. Specifically, as the horizontal quantization ratio $\kappa_{xy}$ increases, the obtained suboptimal path becomes less flexible and results in longer path length, which is due to the more coarse grid/SINR quantization. Particularly, for the case of $\bar{\gamma}_{\mathrm{T}}=3$ dB, the suboptimal paths tend to go back and forth, in contrast to the optimal path which is direct and more efficient since it allows for more flexible turns. Moreover, it is observed that as the expected SINR target increases from $\bar{\gamma}_{\mathrm{T}}=0$ dB to $\bar{\gamma}_{\mathrm{T}}=3$ dB, the feasible grid points become more sparse, as a result of which both the optimal path and suboptimal paths become more complicated.

\begin{figure}[t]
	\centering
	\includegraphics[width=7cm]{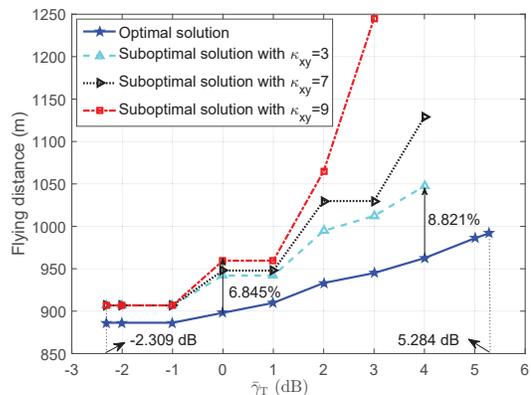}
		\vspace{-4mm}
		\caption{Flying distance versus $\bar{\gamma}_{\mathrm{T}}$ for proposed path solutions with different quantization ratios.}\label{distance}
\end{figure}

Next, we show in Fig. \ref{distance} the required flying distance from $U_0$ to $U_F$ with the proposed optimal and suboptimal solutions versus the expected SINR target $\bar{\gamma}_{\mathrm{T}}$. Note that the minimum value of $[\bar{\mv{S}}]_{i,j,k}$ for all $i,j\in \mathcal{D}$ and $k\in \mathcal{Z}$ on the given map is $-2.309$ dB; moreover, it is found via the BFS algorithm that (P1) becomes infeasible if $\bar{\gamma}_{\mathrm{T}}>5.284$ dB for this setup. Thus, the range of $\bar{\gamma}_{\mathrm{T}}$ in Fig. \ref{distance} is set as $[-2.309,5.284]$ dB. It is observed from Fig. \ref{distance} that the optimal solution is feasible for all values of $\bar{\gamma}_{\mathrm{T}}$, while the suboptimal solution with $\kappa_{xy}=9$ becomes infeasible when $\bar{\gamma}_{\mathrm{T}}>3$ dB, and the suboptimal solutions with $\kappa_{xy}=7$ and $\kappa_{xy}=3$ become infeasible when $\bar{\gamma}_{\mathrm{T}}>4$ dB. Moreover, the required flying distance for the suboptimal solution increases as $\kappa_{xy}$ increases, since the increasingly coarse grid/SINR quantization yields less flexibility in the path design, which thus validates the performance-complexity trade-off discussed in Section \ref{sec_sub}. On the other hand, it is worth noting that the performance loss of the suboptimal solutions compared to the optimal solution is generally small, especially in the low-to-medium expected SINR target regime. For example, when $\bar{\gamma}_{\mathrm{T}}\leq 0$ dB, all the suboptimal solutions require at most $6.845\%$ more flying distance than the optimal solution. Moreover, among all expected SINR target values shown in Fig. \ref{distance}, the maximum percentage of additional flying distance required for the suboptimal solution with $\kappa_{xy}=3$ is only $8.821\%$, yet its required complexity is reduced by $88.89\%$ (i.e., $1-\frac{1}{\kappa_{xy}^2}$) compared to the optimal solution. This thus validates the efficacy of our proposed grid quantization method and the corresponding suboptimal designs, by exploiting the spatial channel/SINR correlation among neighboring grid points.

\vspace{-3mm}
\subsection{Effectiveness of Interference-Aware Path Planning}
In this subsection, we examine the effectiveness of the proposed interference-aware path planning based on radio maps. For comparison, we also consider the following benchmark schemes without (full) interference awareness, where path planning is performed based on upper or lower bounds of the loading factors instead of their exact values:
\begin{itemize}[leftmargin=*]
	\item {\bf{Benchmark scheme 1 (Path planning assuming worse-case interference)}}: In this scheme, we perform path planning by considering the worst-case interference, where the SINR map is constructed by replacing the exact loading factor at each $m$th GBS with its upper bound denoted by $l_m^{\max}$. Note that this corresponds to an ``overestimate'' of the interference level at every UAV location.
	\item {\bf{Benchmark scheme 2 (Path planning assuming no interference)}}: In this scheme, we consider path planning without the knowledge of interference distribution and thus assuming zero interference at all locations, where the SINR map becomes the SNR map with zero loading factor for all GBSs. Note that this corresponds to an ``underestimate'' of the interference level at every UAV location.
\end{itemize}
In the following, we compare the above two benchmark schemes with our proposed optimal solution for two different sets of actual loading factors: $\mv{l}=\bar{\mv{l}}$ and $\mv{l}=0.4\bar{\mv{l}}$, which correspond respectively to the loading factor upper bounds $l_m^{\max}=1,\ \forall m\in \mathcal{M}$ and $l_m^{\max}=0.4,\ \forall m\in \mathcal{M}$. 
\begin{figure}[t]
	\centering
	\subfigure[$\mv{l}=\bar{\mv{l}}$]{
		\includegraphics[width=7cm]{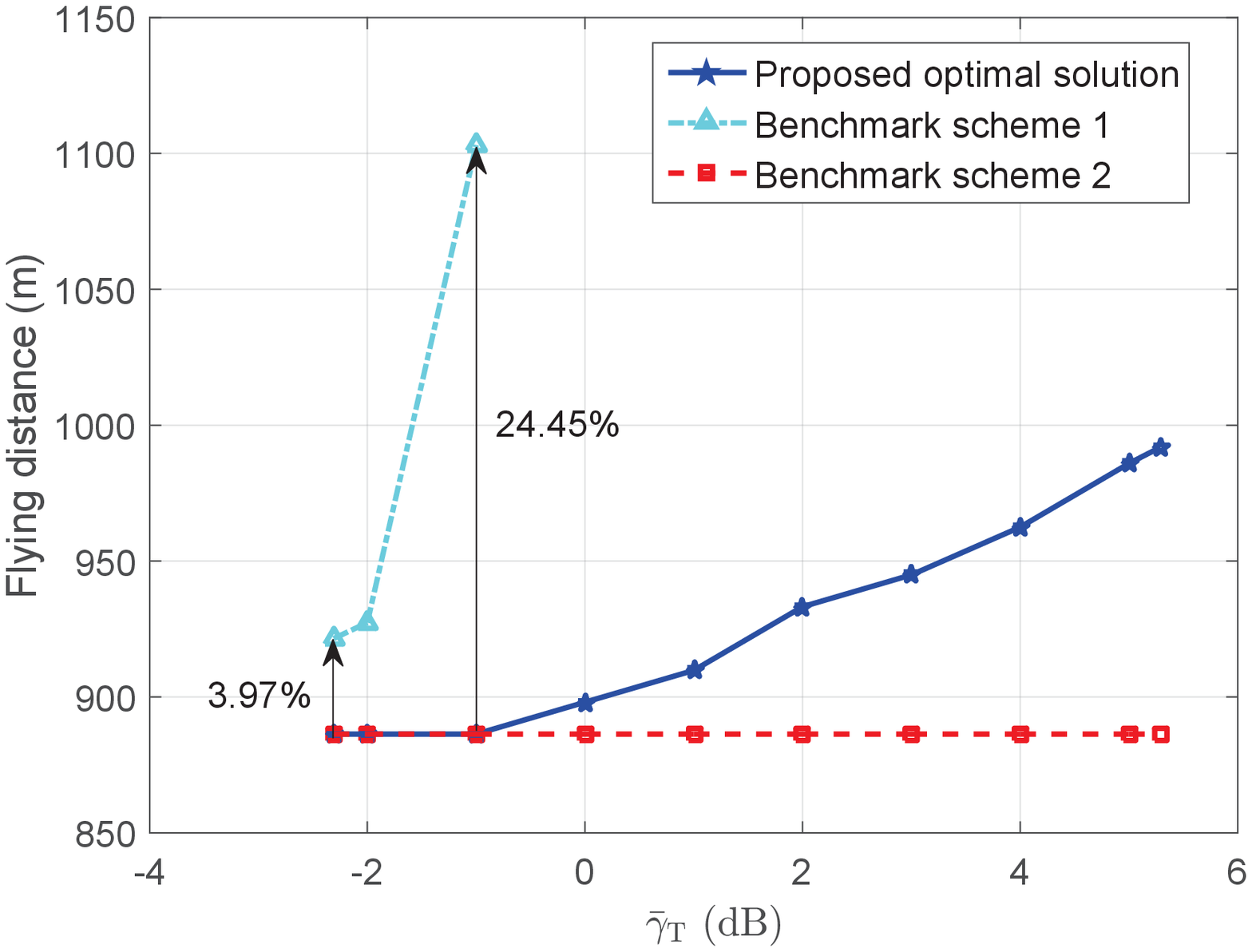}}
	\subfigure[$\mv{l}=0.4\bar{\mv{l}}$]{
		\includegraphics[width=7cm]{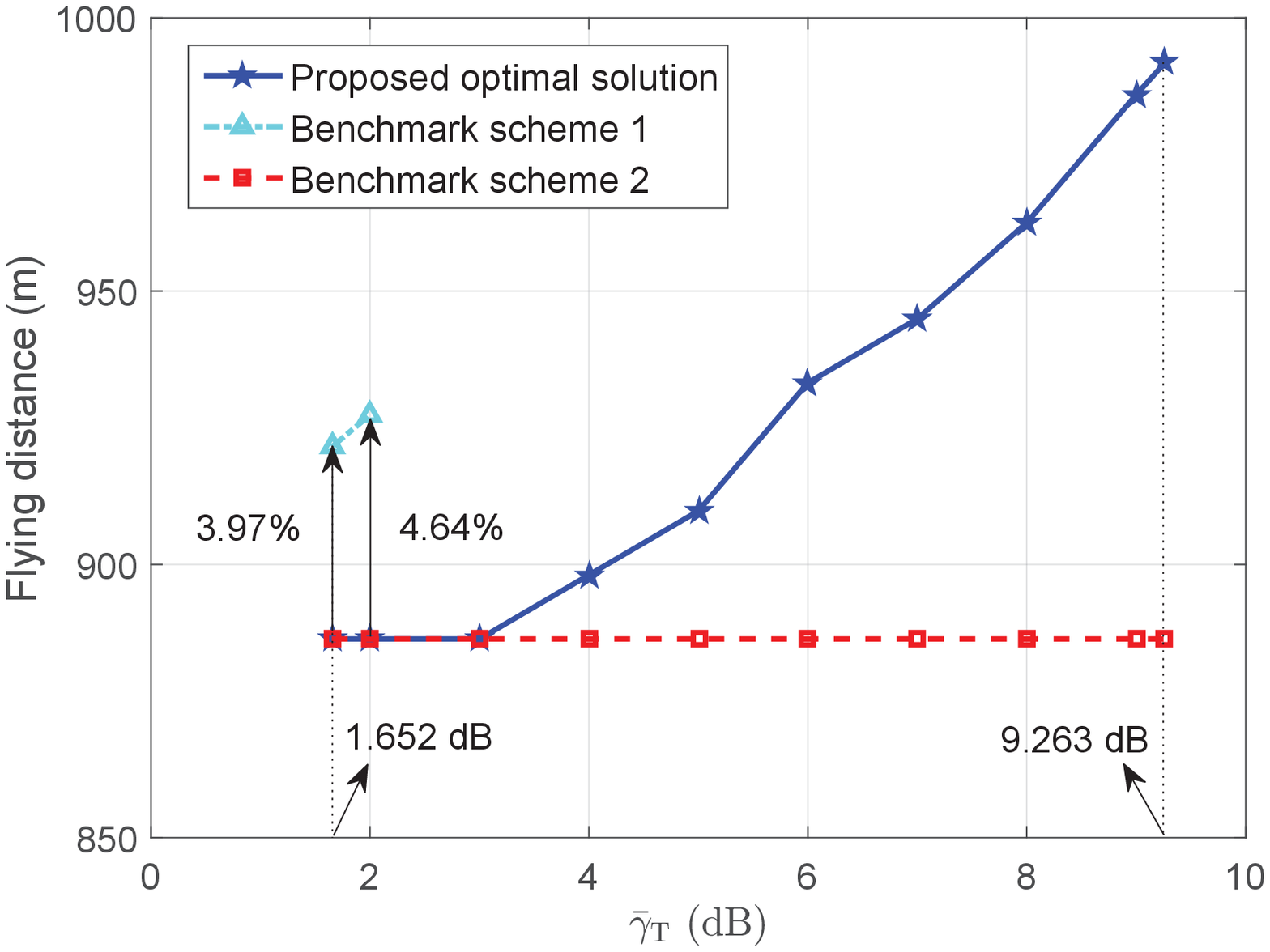}}
	\vspace{-3mm}
	\caption{Flying distance versus $\bar{\gamma}_{\mathrm{T}}$ for the proposed and benchmark scheme path solutions.}\label{Distance}
\end{figure}
\begin{figure}[t]
	\centering
	\subfigure[$\mv{l}=\bar{\mv{l}}$]{
		\includegraphics[width=7cm]{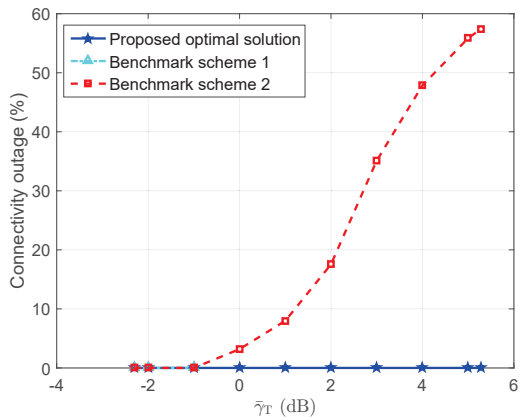}}
	\subfigure[$\mv{l}=0.4\bar{\mv{l}}$]{
		\includegraphics[width=7cm]{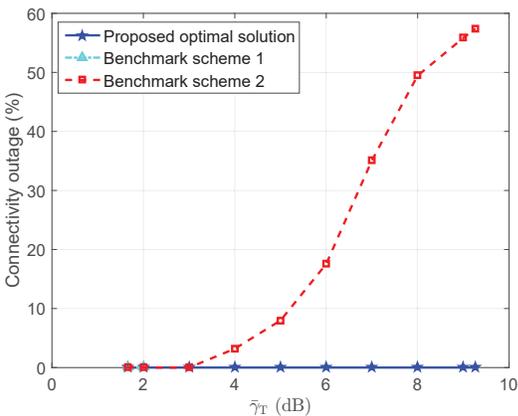}}
	\vspace{-3mm}
	\caption{Connectivity outage versus $\bar{\gamma}_{\mathrm{T}}$ for the proposed and benchmark scheme path solutions.}\label{outage}
\end{figure}

\begin{figure}[t]
	\centering
	\subfigure[$\mv{l}=\bar{\mv{l}}$]{
		\includegraphics[width=9.5cm]{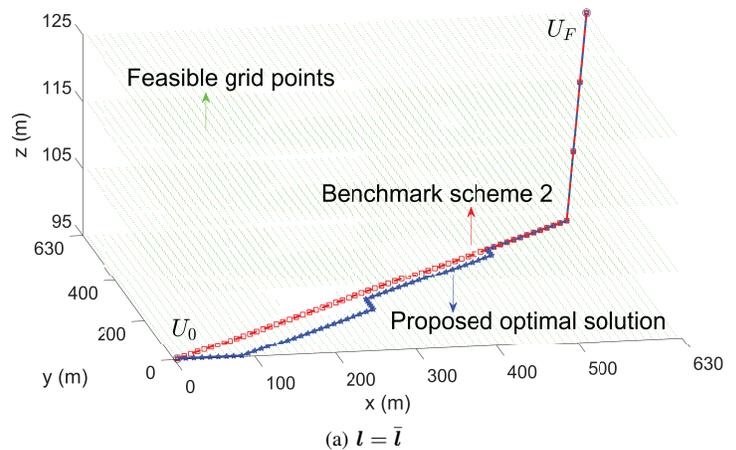}}
	\subfigure[$\mv{l}=0.4\bar{\mv{l}}$]{
		\includegraphics[width=9.5cm]{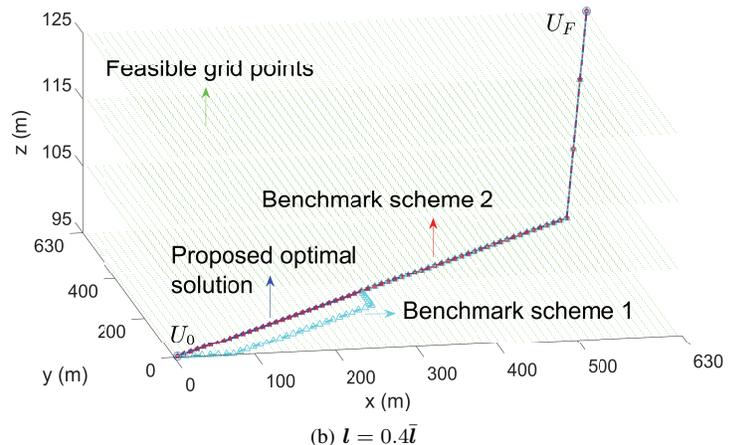}}
	\vspace{-3mm}
	\caption{Illustration of the proposed and benchmark scheme path solutions at $\bar{\gamma}_{\mathrm{T}}=2$ dB.}\label{Interference_Traj}
\end{figure}

First, we show in Fig. \ref{Distance} the required flying distance for these solutions versus the expected SINR target $\bar{\gamma}_{\mathrm{T}}$. It is observed that as the loading factors decrease from $\mv{l}=\bar{\mv{l}}$ to $\mv{l}=0.4\bar{\mv{l}}$, the maximum achievable expected SINR target $\bar{\gamma}_{\mathrm{T}}$ is increased from $5.284$ dB to $9.263$ dB, since the reduced loading factors result in less severe interference at the UAV. Moreover, it is observed that benchmark scheme 1 becomes infeasible when $\bar{\gamma}_{\mathrm{T}}>-1$ dB and $\bar{\gamma}_{\mathrm{T}}>2$ dB for the case of $\mv{l}=\bar{\mv{l}}$ and $\mv{l}=0.4\bar{\mv{l}}$, respectively, due to its overestimation of the interference level. On the other hand, for the case where benchmark scheme 1 is feasible, it requires significantly increased flying distance compared to the proposed solution (e.g., $24.45\%$ and $4.64\%$ at $\bar{\gamma}_{\mathrm{T}}=-1$ dB and $\bar{\gamma}_{\mathrm{T}}=2$ dB for $\mv{l}=\bar{\mv{l}}$ and $\mv{l}=0.4\bar{\mv{l}}$, respectively). On the other hand, although benchmark scheme 2 requires the minimum flying distance at all expected SINR targets, which corresponds to the (quantized) straight flight from $U_0$ to $U_F$, it results in violations of the expected SINR target during the flight. Specifically, we show in Fig. \ref{outage} the \emph{connectivity outage} of the proposed and benchmark schemes, which is defined as the percentage of flying distance that violates the expected SINR target. It is observed from Fig. \ref{outage} that the connectivity outage for our proposed solution and benchmark scheme 1 is always zero; while that for benchmark scheme 2 increases drastically as the expected SINR target increases. This is because with more stringent SINR constraint, benchmark scheme 2 by assuming zero interference tends to result in more failures for meeting this constraint over its flight. Furthermore, we illustrate the proposed path solution and the two benchmark scheme solutions in Fig. \ref{Interference_Traj} under $\bar{\gamma}_{\mathrm{T}}=2$ dB, for both sets of loading factors. It is observed that under the same expected SINR target, the feasible grid points become more sparse as the loading factors increase, since they impose more stringent constraint on the channel gain between the UAV and its associated GBS. It is also observed that for the case of $\mv{l}=\bar{\mv{l}}$, benchmark scheme 1 is infeasible, while benchmark scheme 2 results in substantial connectivity outage; while for the case of $\mv{l}=0.4\bar{\mv{l}}$, benchmark scheme 1 results in larger path length than the proposed solution. The above results therefore validate the effectiveness of our proposed interference-aware path planning solution based on the actual SINR map.

\vspace{-2mm}
\subsection{Effectiveness of Terrain-Aware Path Planning}
Note that a key advantage of the proposed radio map based path planning approach lies in its ability to capture the different channel conditions due to different terrain characteristics over the UAV's fly region (e.g., LoS versus NLoS channels with the GBSs), thereby providing communication performance guarantee at every location during its flight. In this subsection, we evaluate the performance gain of the proposed terrain-aware path planning via comparison with the following benchmark schemes without exploiting the heterogeneous terrain/channel characteristics:
\begin{itemize}[leftmargin=*]
	\item {\bf{Benchmark scheme 3 (LoS channel based path planning)}}: In this scheme, we construct the channel gain maps and the SINR map by assuming that the channel between every UAV location and each GBS follows the LoS model given in (\ref{channel_LoS}). Note that this corresponds to an ``overestimate'' of both the signal power and the interference power, since the LoS-based channel gain is generally larger than the NLoS-based channel gain.
	\item {\bf{Benchmark scheme 4 (NLoS channel based path planning)}}: In this scheme, we construct the channel gain maps and the SINR map by assuming that the channel between every UAV location and each GBS follows the NLoS model given in (\ref{channel_NLoS}). Note that this corresponds to an ``underestimate'' of both the signal power and the interference power.
\end{itemize}

\begin{figure}[t]
	\centering
	\includegraphics[width=7cm]{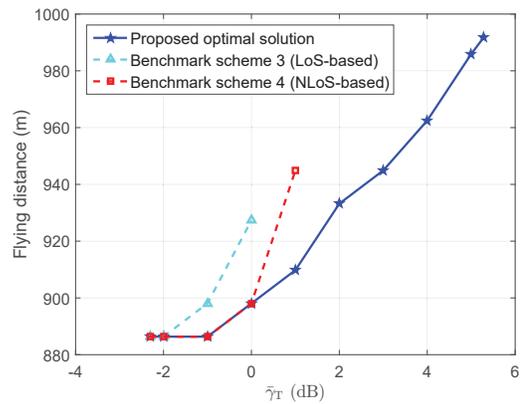}
	\vspace{-4mm}
	\caption{Flying distance versus $\bar{\gamma}_{\mathrm{T}}$ for terrain-aware and terrain-unaware path planning.}\label{LoS_NLoS}
\end{figure}

In Fig. \ref{LoS_NLoS}, we show the flying distance versus the expected SINR target $\bar{\gamma}_{\mathrm{T}}$ for our proposed optimal solution and the above two benchmark schemes, under the loading factor set $\mv{l}=\bar{\mv{l}}$. It is observed that benchmark schemes 3 and 4 (named as terrain-unaware path planning) become infeasible after the fourth and fifth sample points of the expected SINR targets, respectively. Moreover, it is observed that when the two benchmark schemes are feasible, they require significantly increased flying distance compared to the proposed solution (termed as terrain-aware path planning).\footnote{Under this setup, no connectivity outage is observed for the benchmark schemes.} This shows that without accurate knowledge of the terrain-specific channel conditions, assuming LoS or NLoS channels for all locations over the UAV's fly region leads to substantial performance loss due to the inaccurate channel gain maps and SINR map constructed.

\begin{figure*}[t]
	\centering
	\subfigure[GBS-UAV associations over $\mathcal{U}$ at $H=125$ m]{
		\includegraphics[width=7cm]{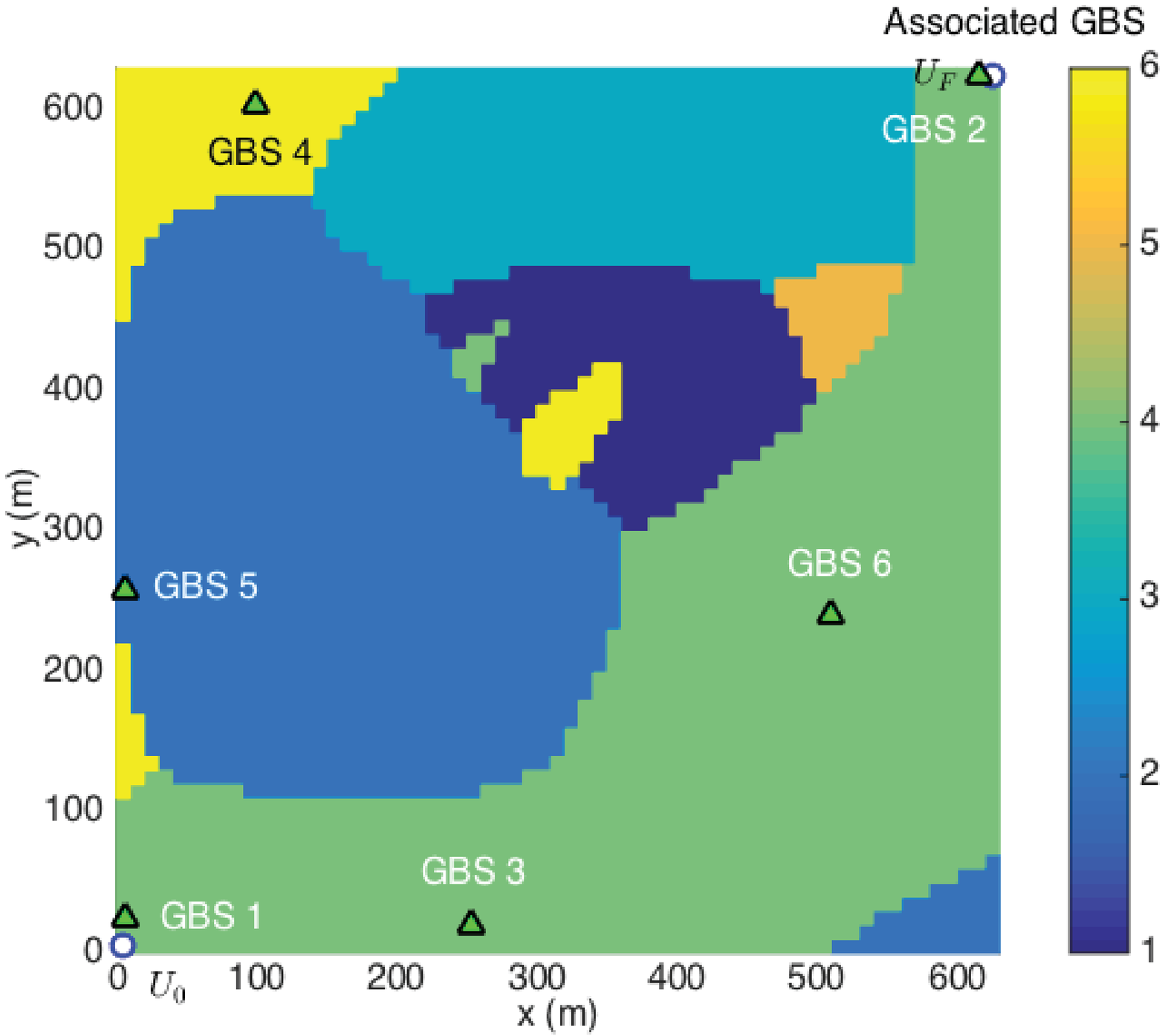}}
	\subfigure[SINR map over $\mathcal{U}$ at $H=125$ m]{
		\includegraphics[width=7cm]{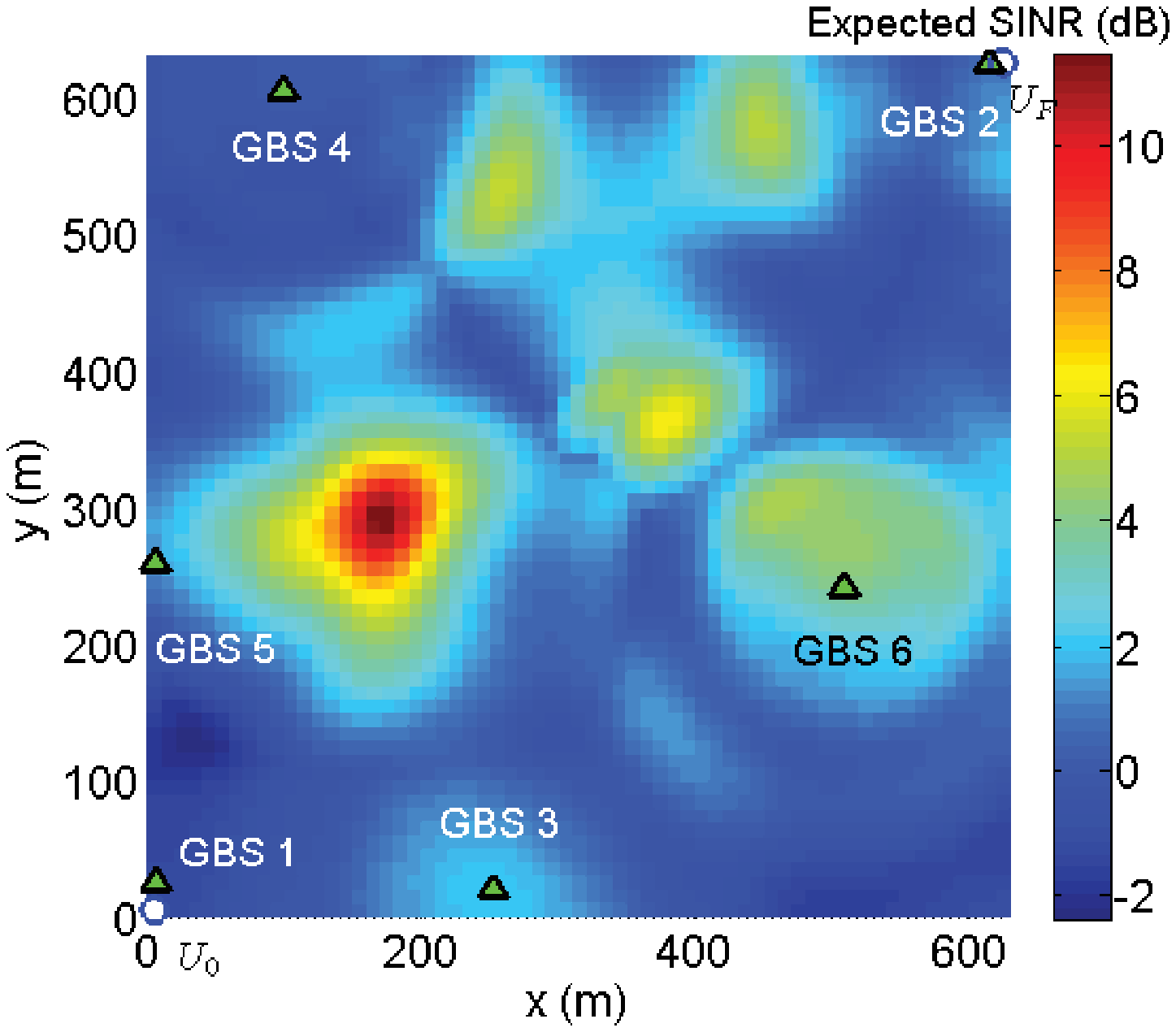}}
	\vspace{-3mm}
	\caption{Illustration of the GBS-UAV associations and the SINR map with downtilted GBS antennas.}\label{Antenna_Map}
	\vspace{-3mm}
\end{figure*}

\vspace{-4mm}
\subsection{Performance of Proposed Solution under Downtilted GBS Antennas}
\begin{figure}[t]
	\centering
	\includegraphics[width=9cm]{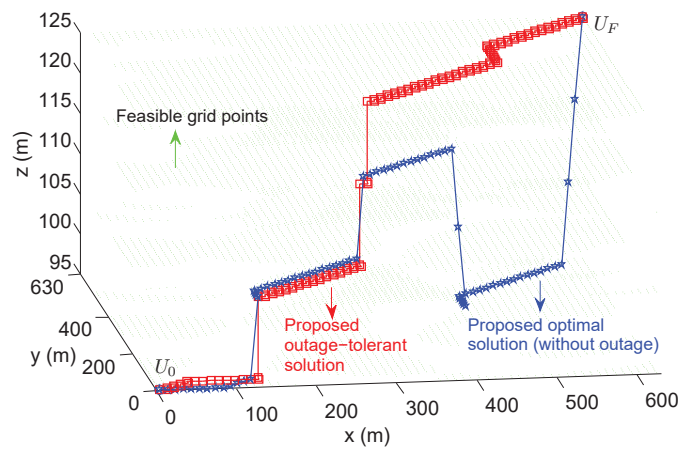}
	\vspace{-5mm}
	\caption{Illustration of the proposed optimal solution and outage-tolerant solution (with $O_T=50$ m) at $\bar{\gamma}_{\mathrm{T}}=-0.2$ dB.}\label{Antenna_Path}
\end{figure}

Finally, we evaluate the performance of our proposed solution under the downtilted GBS antenna model for the current 4G LTE networks. In Fig. \ref{Antenna_Map} (a) and (b), we show the GBS-UAV associations and the SINR map at $H=125$ m, respectively. By comparing Fig. \ref{Antenna_Map} with Fig. \ref{SINR_Map}, it is observed that with downtilted GBS antennas, the SINR map is more heterogeneous compared to the case with fixed and isotropic antenna gain (e.g., the SINR peaks in Fig. \ref{Antenna_Map} (b) are generally not close to the GBSs as in Fig. \ref{SINR_Map} (b)). This is because the UAV in the sky is served by the weak sidelobes of the downtilted GBS antennas in general, which makes the channel gain non-monotonic over the GBS-UAV distance. It is also observed from Fig. \ref{Antenna_Map} (a) that the GBS associated with each UAV location is not the closest one in general. In Fig. \ref{Antenna_Path}, we show the optimal path solution under SINR target $\bar{\gamma}_{\mathrm{T}}=-0.2$ dB. It can be observed that the proposed design is still capable of finding a feasible and shortest path effectively. Moreover, it is worth pointing out that due to the more heterogeneity of the SINR map under the downtilted GBS antennas, there may be more frequent ``coverage holes'' in the sky that cannot meet the SINR target, while avoiding them completely may yield substantially longer flying distance as illustrated in Fig. \ref{Antenna_Path}. To address this issue, we propose a new \emph{``outage-tolerant''} path planning solution by modifying the optimal solution in Section V, where we relax the connectivity constraint by allowing a maximum flying duration with communication outage denoted by $O_T$ (m) during the flight. To incorporate this outage tolerance, we assume that the UAV can fly from a feasible grid point $U_D(i,j,k)$ to another feasible grid point $U_D(i',j',k')$ through the path $U_D(i,j,k)-U_D(i',j,k)-U_D(i',j',k)-U_D(i',j',k')$ (i.e., along the $x$, $y$, and $z$ axes, sequentially) if the maximum path length in outage is no greater than $O_T$. This can be done by constructing  the graph $G_{\mathrm{D}}$ in Section V with additional edges between the aforementioned pairs of feasible grid points whose weights are set as their distances, and then solving the SPP over the graph. In Fig. \ref{Antenna_Path}, we illustrate the proposed outage-tolerant solution under $O_T=50$ m and $\bar{\gamma}_{\mathrm{T}}=-0.2$ dB. It can be observed that the outage-tolerant path is more efficient than the optimal path without outage due to more feasible fly zones, with a flying distance of $933.210$ m that is smaller than $945.924$ m for the case without outage.

\vspace{-3mm}
\section{Concluding Remarks}\label{sec_conclusion}
This paper investigated the interference-aware 3D path planning for a cellular-connected UAV to minimize its flying distance from given initial to final locations, subject to a communication quality constraint specified by an expected receive SINR target with its associated GBSs. We presented a new path optimization framework by utilizing radio maps, which characterize the large-scale channel gains between each GBS and uniformly sampled locations on a 3D grid as well as the expected SINR levels over these sampled 3D locations with optimal GBS association. Based on the SINR map, the optimal path solution was obtained by solving an equivalent SPP, and a suboptimal solution with lower complexity was proposed based on a grid quantization method. Numerical results validated the efficacy of the proposed optimal and suboptimal solutions, and showed their performance-complexity trade-off. The proposed solutions were also shown to achieve significantly improved performance in comparison with various benchmark schemes with imperfect/partial knowledge of the interference and/or channel spatial distribution. 

There are a number of work directions that are worthy of further investigation in the future. First, under the proposed 3D path planning framework based on radio maps, various other utility functions can be considered in the problem objective or constraints such as the general outage cost function proposed in \cite{trajectoryoutage}, the average communication rate over the UAV's entire flight, and the number of \emph{handovers} between different GBSs along the UAV's flight, since frequent handovers will lead to additional delay in GBS-UAV communications \cite{Mobility_Sky,Handover_Challenge}. It is also of high practical interest to extend our results to the more general case with multiple UAVs by considering their collision avoidance constraints (see, e.g., \cite{Chao}). Moreover, we considered downlink communication in this paper for the purpose of exposition, while the interference-aware path planning under uplink communication constraints is also an interesting problem to study, which is different from the downlink case since the interference is caused by the UAV to the non-associated GBSs in the uplink. Furthermore, it is interesting to jointly design path planning with advanced interference mitigation techniques (e.g., dynamic 3D beamforming), which will lead to more complicated radio maps and thus call for new and more efficient map modeling/estimation methods. Finally, it is worthwhile to develop robust path planning solutions under practically imperfect/partial radio map knowledge.

\bibliographystyle{IEEEtran}
\bibliography{TWC_3D_Radio_Map}

\begin{thebibliography}{10}
\providecommand{\url}[1]{#1}
\csname url@samestyle\endcsname
\providecommand{\newblock}{\relax}
\providecommand{\bibinfo}[2]{#2}
\providecommand{\BIBentrySTDinterwordspacing}{\spaceskip=0pt\relax}
\providecommand{\BIBentryALTinterwordstretchfactor}{4}
\providecommand{\BIBentryALTinterwordspacing}{\spaceskip=\fontdimen2\font plus
\BIBentryALTinterwordstretchfactor\fontdimen3\font minus
  \fontdimen4\font\relax}
\providecommand{\BIBforeignlanguage}[2]{{%
\expandafter\ifx\csname l@#1\endcsname\relax
\typeout{** WARNING: IEEEtran.bst: No hyphenation pattern has been}%
\typeout{** loaded for the language `#1'. Using the pattern for}%
\typeout{** the default language instead.}%
\else
\language=\csname l@#1\endcsname
\fi
#2}}
\providecommand{\BIBdecl}{\relax}
\BIBdecl

\bibitem{GC}
S.~Zhang and R.~Zhang, ``{Radio map based path planning for cellular-connected
  UAV},'' in \emph{Proc. IEEE Global Commun. Conf. (Globecom)}, Dec. 2019.

\bibitem{cellularUAV}
S.~Zhang, Y.~Zeng, and R.~Zhang, ``{Cellular-enabled UAV communication: A
  connectivity-constrained trajectory optimization perspective},'' \emph{IEEE
  Trans. Commun.}, vol.~67, no.~3, pp. 2580--2604, Mar. 2019.

\bibitem{LTE_Sky}
B.~V.~D. Bergh, A.~Chiumento, and S.~Pollin, ``{LTE in the sky: Trading off
  propagation benefits with interference costs for aerial nodes},'' \emph{IEEE
  Commun. Mag.}, vol.~54, no.~5, pp. 44--50, May 2016.

\bibitem{LinSky}
X.~Lin \emph{et~al.}, ``{The sky is not the limit: LTE for unmanned aerial
  vehicles},'' \emph{IEEE Commun. Mag.}, vol.~56, no.~4, pp. 204--210, Apr.
  2018.

\bibitem{Potential}
Y.~Zeng, J.~Lyu, and R.~Zhang, ``{Cellular-connected UAV: Potential,
  challenges, and promising technologies},'' \emph{IEEE Wireless Commun.},
  vol.~26, no.~1, pp. 120--127, Feb. 2019.

\bibitem{Network_Connected}
J.~Lyu and R.~Zhang, ``{Network-connected UAV: 3D system modeling and coverage
  performance analysis},'' \emph{IEEE IoT J.}, vol.~6, no.~4, pp. 7048--7060,
  Aug. 2019.

\bibitem{multibeam}
L.~Liu, S.~Zhang, and R.~Zhang, ``{Multi-beam UAV communication in cellular
  uplink: Cooperative interference cancellation and sum-rate maximization},''
  \emph{IEEE Trans. Wireless Commun.}, vol.~18, no.~10, pp. 4679--4691, Oct.
  2019.

\bibitem{NOMA}
L.~Liu, S.~Zhang, and R.~Zhang, ``{Exploiting NOMA for multi-beam UAV
  communication in cellular uplink},'' in \emph{Proc. IEEE Int. Conf. Commun.
  (ICC)}, May 2019.

\bibitem{Mei2}
W.~Mei and R.~Zhang, ``{Uplink cooperative NOMA for cellular-connected UAV},''
  \emph{IEEE J. Sel. Topics Signal Process.}, vol.~13, no.~3, pp. 644--656,
  Jun. 2019.

\bibitem{Mei}
W.~Mei, Q.~Wu, and R.~Zhang, ``{Cellular-connected UAV: Uplink association,
  power control and interference coordination},'' \emph{IEEE Trans. Wireless
  Commun.}, vol.~18, no.~11, pp. 5380--5393, Nov. 2019.

\bibitem{Cognitive}
Y.~Huang, W.~Mei, J.~Xu, L.~Qiu, and R.~Zhang, ``{Cognitive UAV communication
  via joint maneuver and power control},'' \emph{IEEE Trans. Commun.}, vol.~67,
  no.~11, pp. 7872--7888, Nov. 2019.

\bibitem{Howto}
H.~C. Nguyen \emph{et~al.}, ``{How to ensure reliable connectivity for aerial
  vehicles over cellular networks},'' \emph{IEEE Access}, vol.~6, pp.
  12\,304--12\,317, Feb. 2018.

\bibitem{trajectoryoutage}
S.~Zhang and R.~Zhang, ``{Trajectory optimization for cellular-connected UAV
  under outage duration constraint},'' \emph{J. Commun. Inf. Network.}, vol.~4,
  no.~4, pp. 55--71, Dec. 2019.

\bibitem{trajectoryoutageICC}
S.~Zhang and R.~Zhang, ``{Trajectory design for cellular-connected UAV under
  outage duration constraint},'' in \emph{Proc. IEEE Int. Conf. Commun. (ICC)},
  May 2019.

\bibitem{Disconnectivity}
E.~Bulut and I.~Guvenc, ``{Trajectory optimization for cellular-connected UAVs
  with disconnectivity constraint},'' in \emph{Proc. IEEE Int. Conf. Commun.
  (ICC) Wkshps.}, May 2018.

\bibitem{Power_Efficient}
B.~Khamidehi and E.~S. Sousa, ``{Power-efficient trajectory optimization for
  the cellular-connected aerial vehicles},'' in \emph{Proc. IEEE Int. Symp.
  Personal, Indoor, Mobile Radio Commun. (PIMRC)}, Sep. 2019.

\bibitem{3DMap}
O.~Esrafilian, R.~Gangula, and D.~Gesbert, ``{3D-map assisted UAV trajectory
  design under cellular connectivity constraints},'' in \emph{Proc. IEEE Int.
  Conf. Commun. (ICC)}, May 2020.

\bibitem{Interference_YW}
X.~Mu, Y.~Liu, L.~Guo, C.~Dong, and J.~Lin, ``{Interference-aware trajectory
  design for ground-aerial uplink NOMA cellular networks},'' in \emph{Proc.
  IEEE Global Commun. Conf. (Globecom)}, Dec. 2019.

\bibitem{Reinforcement}
U.~Challita, W.~Saad, and C.~Bettstetter, ``{Interference management for
  cellular-connected UAVs: A deep reinforcement learning approach},''
  \emph{IEEE Trans. Wireless Commun.}, vol.~18, no.~4, pp. 2125--2140, Apr.
  2019.

\bibitem{URLLCPollin}
M.~M. Azari, F.~Rosas, K.-C. Chen, and S.~Pollin, ``{Ultra reliable UAV
  communication using altitude and cooperation diversity},'' \emph{IEEE Trans.
  Commun.}, vol.~66, no.~1, pp. 330--344, Jan. 2018.

\bibitem{LAP}
A.~Al-Hourani, S.~Kandeepan, and S.~Lardner, ``{Optimal LAP altitude for
  maximum coverage},'' \emph{IEEE Wireless Commun. Lett.}, vol.~3, no.~6, pp.
  569--572, Dec. 2014.

\bibitem{CoMP}
L.~Liu, S.~Zhang, and R.~Zhang, ``{CoMP in the sky: UAV placement and movement
  optimization for multi-user communications},'' \emph{IEEE Trans. Commun.},
  vol.~67, no.~8, pp. 5645--5658, Aug. 2019.

\bibitem{You}
C.~You and R.~Zhang, ``{3D trajectory optimization in {Rician} fading for
  UAV-enabled data harvesting},'' \emph{IEEE Trans. Wireless Commun.}, vol.~18,
  no.~6, pp. 3192--3207, Jun. 2019.

\bibitem{Map_based}
O.~Esrafilian, R.~Gangula, and D.~Gesbert, ``{Learning to communicate in
  UAV-aided wireless networks: Map-based approaches},'' \emph{IEEE IoT J.},
  vol.~6, no.~2, pp. 1791--1802, Apr. 2018.

\bibitem{YW}
X.~Liu, Y.~Liu, Y.~Chen, and L.~Hanzo, ``{Trajectory design and power control
  for multi-UAV assisted wireless networks: A machine learning approach},''
  \emph{IEEE Trans. Veh. Technol.}, vol.~68, no.~8, pp. 7957--7969, Aug. 2019.

\bibitem{Hybrid}
C.~You and R.~Zhang, ``{Hybrid offline-online design for UAV-enabled data
  harvesting in probabilistic LoS channel},'' \emph{IEEE Trans. Wireless
  Commun.}, vol.~19, no.~6, pp. 3753--3768, Jun. 2020.

\bibitem{engineering}
S.~Bi, J.~Lyu, Z.~Ding, and R.~Zhang, ``{Engineering radio map for wireless
  resource management},'' \emph{IEEE Wireless Commun.}, vol.~26, no.~2, pp.
  133--141, Apr. 2019.

\bibitem{Learning}
J.~Chen, U.~Yatnalli, and D.~Gesbert, ``{Learning radio maps for UAV-aided
  wireless networks: A segmented regression approach},'' in \emph{Proc. IEEE
  Int. Conf. Commun. (ICC)}, May 2017.

\bibitem{Efficient}
J.~Chen, O.~Esrafilian, D.~Gesbert, and U.~Mitra, ``{Efficient algorithms for
  air-to-ground channel reconstruction in UAV-aided communications},'' in
  \emph{Proc. IEEE Global Commun. Conf. (Globecom) Wkshps.}, Dec. 2017.

\bibitem{graph}
D.~B. West, \emph{{Introduction to Graph Theory}}.\hskip 1em plus 0.5em minus
  0.4em\relax Prentice Hall, 2001.

\bibitem{Placement}
J.~Chen and D.~Gesbert, ``Efficient local map search algorithms for the
  placement of flying relays,'' \emph{IEEE Trans. Wireless Commun.}, vol.~19,
  no.~2, pp. 1305--1319, Feb. 2020.

\bibitem{coverage}
H.~Yang, J.~Zhang, S.~H. Song, and K.~B. Letaief, ``{Connectivity-aware UAV
  path planning with aerial coverage maps},'' in \emph{Proc. IEEE Wireless
  Commun. Network. Conf. (WCNC)}, Apr. 2019.

\bibitem{3GPPUAV}
{3GPP TR 36.777}, ``Enhanced {LTE} support for aerial vehicles (release 15),''
  {V15.0.0}.

\bibitem{3GPPmodel}
{3GPP TR 36.873}, ``{Study on 3D channel model for LTE (release 12)},''
  {V12.7.0}.

\bibitem{Mobility_Sky}
R.~Amer, W.~Saad, and N.~Marchetti, ``{Mobility in the sky: Performance and
  mobility analysis for cellular-connected UAVs},'' \emph{IEEE Trans. Commun.},
  vol.~68, no.~5, pp. 3229--3246, May 2020.

\bibitem{Handover_Challenge}
A.~Fakhreddine, C.~Bettstetter, S.~Hayat, R.~Muzaffar, and D.~Emini,
  ``{Handover challenges for cellular-connected drones},'' in \emph{Proc. 5th
  Wkshps. Micro Aerial Veh. Networks, Systems, and Appl.}, Jun. 2019.

\bibitem{Chao}
C.~Shen, T.-H. Chang, J.~Gong, Y.~Zeng, and R.~Zhang, ``{Multi-UAV interference
  coordination via joint trajectory and power control},'' \emph{IEEE Trans.
  Signal Process.}, vol.~68, pp. 843--858, Jan. 2020.

\end{thebibliography}
\end{document}